# Single-Atom Adsorption on h-BN along the Periodic Table of Elements: From Pristine Surface to Vacancy-Engineered Sites

Ana S. Dobrota[1], Natalia V. Skorodumova[2], Igor A. Pašti[1,3]*
[1] *University of Belgrade – Faculty of Physical Chemistry, Studentski trg 12-16, 11000 Belgrade, Serbia*
[2] Applied Physics, Division of Materials Science, Department of Engineering Sciences and Mathematics, Luleå University of Technology, 971 87 Luleå, Sweden
[3] Serbian Academy of Sciences and Arts, Kneza Mihaila 35, 11000 Belgrade, Serbia

*corresponding author
Prof. Igor A. Pašti
*University of Belgrade – Faculty of Physical Chemistry*
*Studentski trg 12-16, 11158 Belgrade, Serbia*
*E-mail:* igor@ffh.bg.ac.rs
*Phone: +381 11 3336 625*

## Abstract

The adsorption of single atoms on pristine and defected hexagonal boron nitride (h-BN) was systematically investigated using density functional theory. Elements from the first three rows of the periodic table, together with selected transition and coinage metals, were examined on the pristine surface and at boron- and nitrogen-vacancy sites. On pristine h-BN, adsorption is generally weak and dominated by dispersion forces, with measurable chemisorption limited to highly electronegative atoms such as C, O, and F. The introduction of vacancies transforms h-BN into a chemically active material, increasing adsorption energies by one to two orders of magnitude. The boron vacancy strongly stabilizes metallic and electropositive species through coordination to undercoordinated nitrogen atoms, whereas the nitrogen vacancy selectively binds electronegative and covalent adsorbates. Scaling of adsorption energies with elemental cohesive energies distinguishes regimes of physisorption, chemisorption, and substitutional stabilization. These insights provide a unified description of adsorption trends across the periodic table and establish defect engineering as an effective strategy for tailoring the catalytic, sensing, and electronic properties of h-BN.

## Graphical abstract

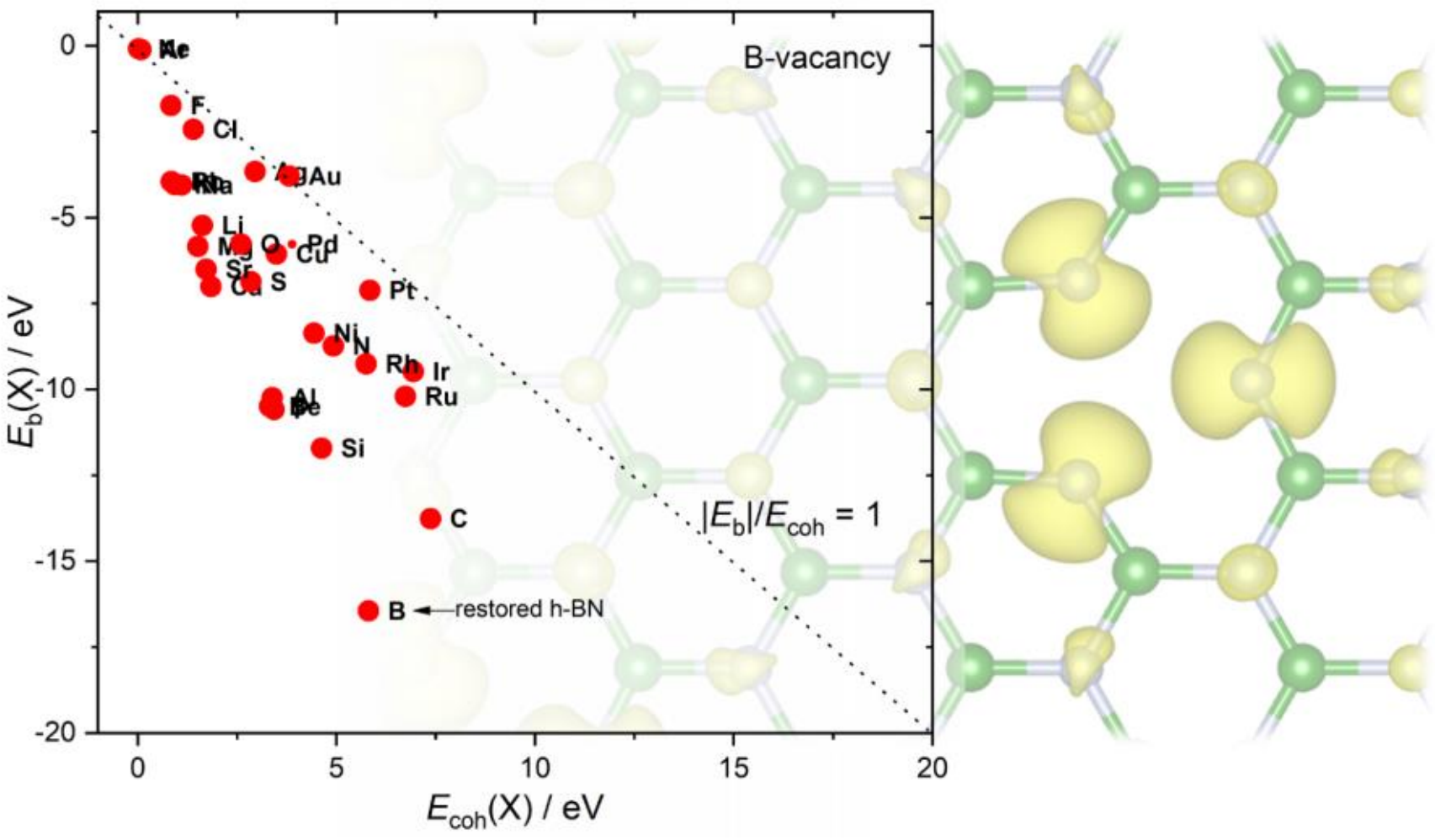

## 1. Introduction

Hexagonal boron nitride (h-BN) is a two-dimensional (2D) layered material with a honeycomb lattice analogous to graphene but distinguished by its wide band gap (~5−6 eV), high thermal and chemical stability, and mechanical robustness, which collectively underpin applications ranging from dielectric substrates in nanoelectronics to protective coatings and heterogeneous catalysis [1]. While the intrinsic chemical inertness of pristine h-BN is desirable for insulating and barrier roles, it limits strong interactions with adsorbates, which is a central requirement for catalysis, molecular recognition, and chemical functionalization [1,2]. As a result, defect engineering has emerged as a key strategy to activate h-BN: single-atom vacancies ($V_B$, $V_N$), divacancies, and substitutional doping introduce under-coordinated sites and localized states that can dramatically alter adsorption energetics and electronic structure [2–4].

A rapidly growing body of density-functional-theory (DFT) literature demonstrates that vacancy sites in h-BN serve as anchoring centers for atoms and small molecules, often converting weak physisorption on the pristine lattice into strong chemisorption accompanied by charge transfer, spin polarization, and pronounced changes in the density of states [2–6]. For example, adsorption of first- and second-row elements (H, Li, C, O, Al, Si, P, S) shows substantially enhanced binding at $V_B$ and $V_N$ relative to the basal plane, together with vacancy-dependent preferences [2]. Anchored transition metals at vacancy sites have been shown to act as single-atom catalysts (SACs) for reactions such as CO oxidation and alkane dehydrogenation [3,4], while decoration of h-BN with heteroatoms has been proposed for functionalization and sensing, where modified adsorption and charge-transfer patterns enhance sensitivity toward $NO_x$ and $CO_x$ molecules [5,6]. Despite this progress, the literature remains fragmented. Most studies address specific chemical families, such as main-group elements [2], selected transition metals for SAC applications [3,4,7], or noble gases on boron-rich BN [8], and often consider either pristine h-BN or a single vacancy type. In contrast to graphene, where systematic studies exist for pristine [9,10] and defective surfaces [11,12], in the case of h-BN, we did not find any systematic study that jointly addresses metals, non-metals, halogens, and noble gases across both $V_B$ and $V_N$ defects within one consistent DFT framework [3–8,13,14].

Here, we address this gap by systematically studying the adsorption of all elements in rows 1−3 of the periodic table, together with catalytically important transition metals (Ni, Ru, Rh, Pd, Ir, Pt) and coinage metals (Cu, Ag, Au), as well as K and Ca. Using a unified DFT framework, we compare adsorption on pristine h-BN with that at boron and nitrogen vacancies to reveal how defect type and adsorbate chemistry govern binding strength and electronic response. This comprehensive dataset captures dispersion-, ionic-, and covalent-binding regimes and exposes clear periodic trends and site selectivity between $V_B$ and $V_N$. By integrating metals, non-metals, halogens, and noble gases within a single consistent approach, this work bridges previously fragmented studies and establishes a quantitative foundation for rational design and defect-driven functionalization of h-BN in catalysis, sensing, and electronic applications, while also presenting a strong basis for training ML-based potentials for rapid screening and materials discovery.

## 2. Computational Methods

We calculated the adsorption of all elements of the periodic table located in rows 1 to 3, in addition to K, Ca, selected d-elements (Ni, Ru, Rh, Pd, Ir, Pt), and coinage metals (Cu, Ag, and Au). Adsorption was modeled on a 4×4 supercell of monolayer h-BN (32 atoms), either pristine or containing a single B- or N-vacancy. The chosen cell size ensures that periodic images of adsorbates and defects remain well separated, thereby avoiding artificial interactions arising from periodic boundary conditions. A vacuum region of 20 Å was added along the direction perpendicular to the surface to prevent spurious interactions between adjacent layers.

First-principles density functional theory (DFT) calculations were carried out using the Vienna Ab initio Simulation Package (VASP) [15–18]. The generalized gradient approximation (GGA) in Perdew, Burke, and Ernzerhof (PBE) parametrization [19] was employed, in conjunction with the projector augmented-wave (PAW) method [20,21]. Standard PAW potentials supplied with the VASP distribution were used throughout. Dispersion interactions were included using the DFT-D3 correction of Grimme [22]. The plane-wave cutoff energy was set to 520 eV, and Gaussian smearing with a width of $\sigma = 0.025$ eV was applied to the occupation of electronic states. Brillouin zone sampling was performed using a Monkhorst–Pack Γ-centered 10×10×1 k-point mesh. Spin polarization was included in all calculations. Structural optimizations were performed until the Hellmann–Feynman forces on all atoms were smaller than $10^{-2}$ eV Å$^{-1}$, corresponding to a total energy convergence of below 0.01 meV.

For adsorption, multiple initial adsorption sites were tested systematically on pristine h-BN (see Supplementary Information, **Figure S1**). In cases of vacancy, we focused on the reactivity of the vacancy sites themselves. The binding energy of an adatom A on h-BN was calculated as:

$$E_b(\mathrm{A}) = E_{\text{h-BN+A}} - E_{\text{h-BN}} - E_{\mathrm{A}} \quad (1)$$

where $E_{\text{h-BN+A}}$ is the total energy of the h-BN slab (pristine or with a vacancy) with adsorbed atom A, $E_{\text{h-BN}}$ is the energy of the clean slab (pristine or defective), and $E_{\mathrm{A}}$ is the energy of the isolated adatom A. Negative values of $E_b(\mathrm{A})$ indicate exothermic adsorption. Visualization of optimized structures was carried out using VESTA [23], while graphical presentation of densities of states (DOS) was performed with the sumo toolkit for VASP [24]. Band gaps were extracted using VASP Kit tools [25]. The charge transfer was analyzed using the Bader algorithm [26] on a charge density grid by Henkelman et al. [27].

## 3. Results and Discussion

### 3.1 Structural models and adsorption sites

Pristine h-BN, as modelled in this work, is a non-magnetic insulator with a band gap of 4.70 eV (**Figure 1**), in line with other theoretical work available in the literature [28]. The formation of nitrogen vacancy ($V_N$) induces defect states and magnetism (total 1 $\mu_B$), which are dominantly located on the boron atoms surrounding the vacancy, and partially on the nitrogen atoms in the second coordination shell of the vacancy site (**Figure 1**, middle, and **Figure 2**, left). The band gap is significantly reduced to 0.51 eV, while the boron atoms around the vacancy share approx. 1 extra electron compared to other B atoms in the layer. Similarly, the formation of a boron vacancy ($V_B$) induces empty states well localized on nitrogen atoms surrounding the vacancy (**Figure 1**, right, and **Figure 2**, right). The system is magnetic with a total magnetization of 3 $\mu_B$, while the band gap is only 0.11 eV.

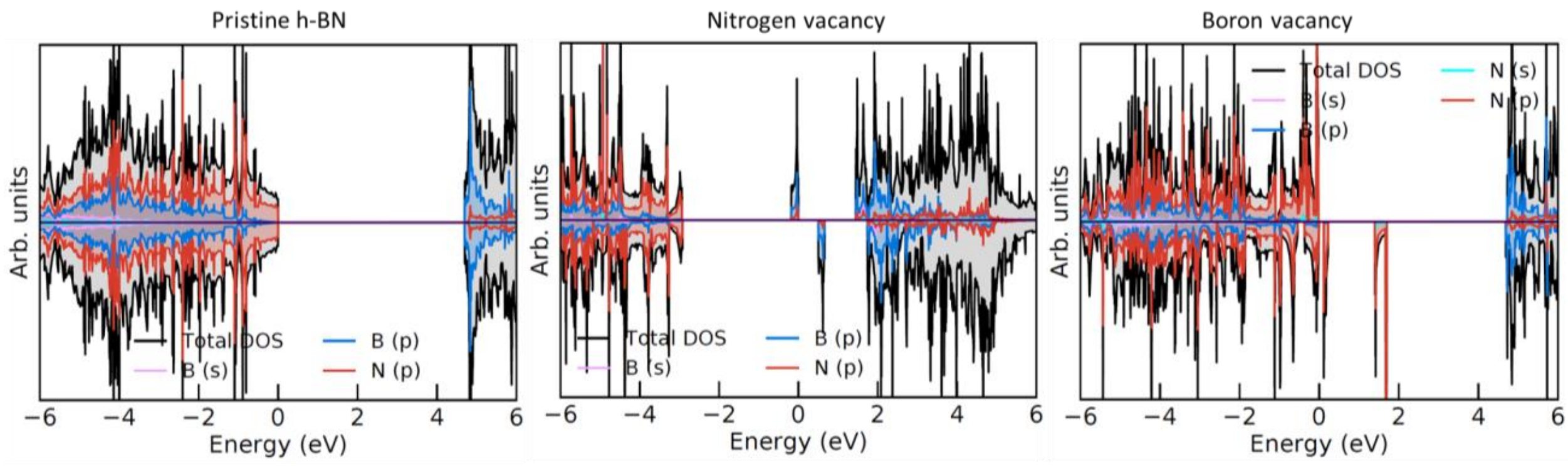

Figure 1. Density of states of pristine h-BN (left), h-BN with nitrogen vacancy (middle), and h-BN with boron vacancy (right).

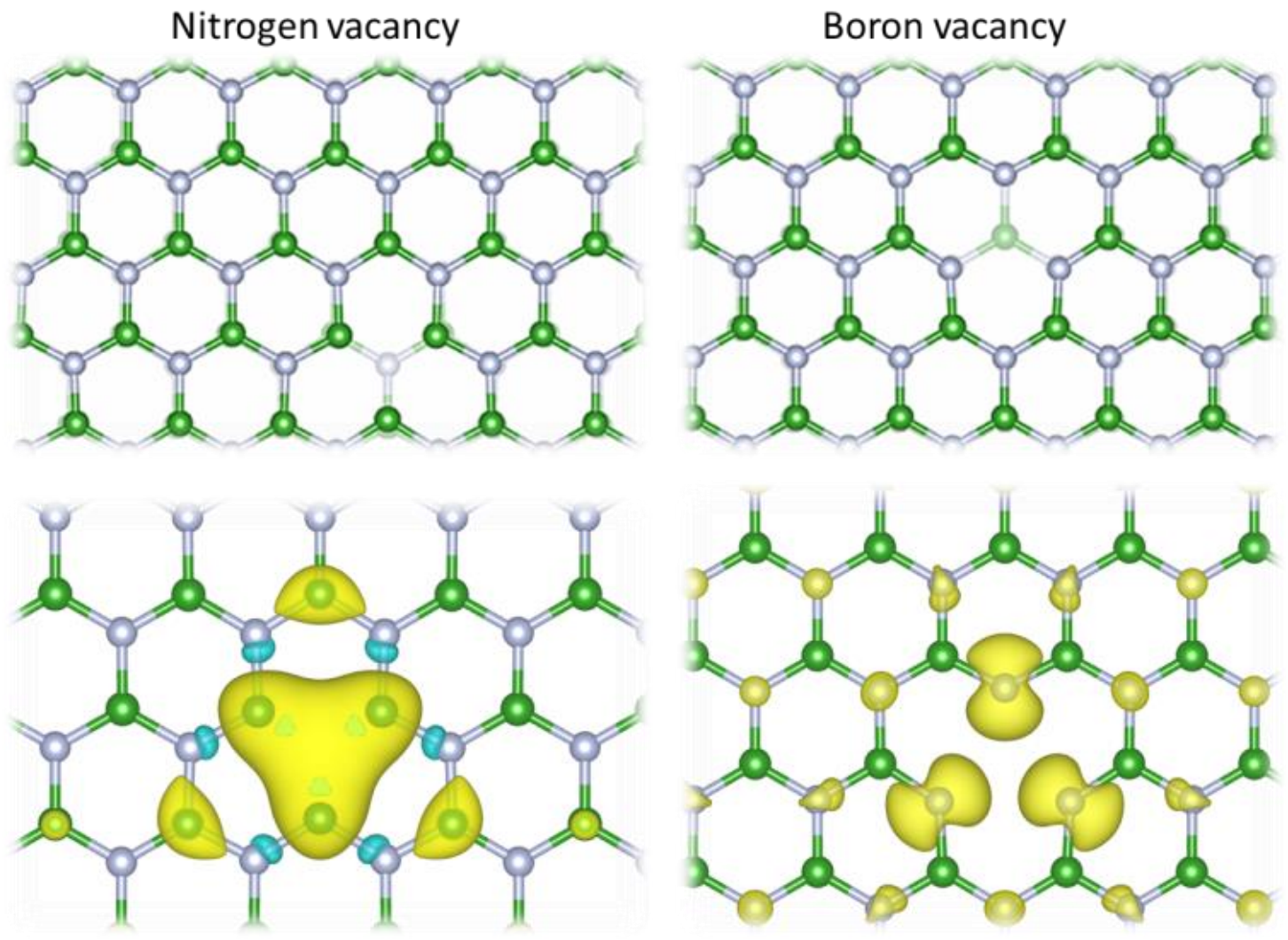

Figure 2. Models of h-BN with nitrogen vacancy (left) and boron vacancy (right). Top row: the vacant structures are overlapped with the pristine h-BN structures, where relaxation of B atoms can be observed towards the vacant N site, while outward relaxation on N atoms from the vacant B site can be observed in the case of $V_B$. Bottom row: magnetization densities of vacant h-BN.

### 3.2 Adsorption on pristine h-BN

The calculated binding energies at the preferred adsorption sites on pristine h-BN, along with magnetizations, charge transfer, and band gaps, are presented in **Table 1**. A complete set of data for different adsorption sites is provided in Table S1 of the Supplementary Information.

Table 1. Preferential adsorption sites, binding energies ($E_b$(X)), total magnetizations ($M_{tot}$), charge transferred from X adatom to the substrate ($\Delta q$(X)), band gaps ($E_{gap}$), nearest neighbors and adsorbate-nearest neighbor distance ($d$) for atomic adsorption onto h-BN.

| PTE part | X | Ads. site | $E_b$(X) / eV | $M_{tot}$ / $\mu_B$ | $\Delta q$(X) / e | $E_{gap}$ / eV | Nearest neighbor | $d$ / Å |
|---|---|---|---|---|---|---|---|---|
| group Ia | H | bridge | −0.05 | 1.00 | 0.00 | 4.13 | B | 2.88 |
| | Li | bridge | −0.31 | 1.00 | 0.16 | 1.02 | N | 3.06 |
| | Na | bridge | −0.21 | 1.00 | 0.15 | 0.77 | N | 3.48 |
| | K | bridge | −0.21 | 1.00 | 0.13 | 0.48 | B | 4.03 |
| | Rb | B–top | −0.21 | 1.00 | 0.13 | 0.31 | B | 4.16 |
| group IIa | Be | B–top | −0.20 | 0.00 | 0.10 | 3.38 | B | 3.05 |
| | Mg | hollow | −0.25 | 0.00 | −0.06 | 3.00 | B | 3.50 |
| | Ca | N–top | −0.31 | 0.00 | 0.27 | 1.68 | N | 3.54 |
| | Sr | hollow | −0.33 | 0.00 | 0.18 | 1.84 | B | 4.04 |
| p – row 2 | B | bridge | −0.85 | 1.00 | −0.45 | 1.22 | N | 1.68 |
| | C | N–top | −1.31 | 2.00 | −0.27 | 1.53 | N | 1.61 |
| | N | bridge | −0.57 | 1.00 | −0.79 | 1.32 | B | 1.52 |
| | O | bridge | −2.08 | 0.00 | −0.90 | 3.34 | B | 1.47 |
| | F | B–top | −1.94 | 0.25 | −0.80 | 0.02 | B | 1.45 |
| p – row 3 | Al | bridge | −0.27 | 1.00 | 0.16 | 0.31 | N | 2.72 |
| | Si | N–top | −0.94 | 2.00 | 0.15 | 1.38 | N | 2.20 |
| | P | B–top | −0.11 | 3.00 | 0.09 | 2.19 | B | 3.24 |
| | S | N–top | −1.06 | 0.00 | −0.75 | 2.97 | N | 1.90 |
| | Cl | N–top | −0.72 | 1.00 | −0.46 | 0.52 | N | 2.59 |
| noble | He | bridge | −0.02 | 0.00 | −0.01 | 4.69 | B | 2.99 |
| | Ne | B–top | −0.04 | 0.00 | −0.01 | 4.67 | B | 3.07 |
| | Ar | hollow | −0.08 | 0.00 | −0.01 | 4.67 | B | 3.57 |
| d – row 4 | Ni | N–top | −1.57 | 0.00 | 0.19 | 0.94 | N | 1.86 |
| | Cu | N–top | −0.39 | 1.00 | −0.49 | 0.87 | N | 2.18 |
| d – row 5 | Ru | hollow | −2.19 | 2.00 | 0.11 | 1.04 | B | 2.27 |
| | Rh | bridge | −1.65 | 1.00 | −0.29 | 0.23 | B | 2.16 |
| | Pd | bridge | −1.12 | 0.00 | −0.79 | 1.59 | N | 2.22 |
| | Ag | bridge | −0.31 | −1.00 | 0.38 | 0.68 | B | 3.02 |
| d – row 6 | Ir | bridge | −1.30 | 1.00 | −1.68 | 0.50 | N | 2.06 |
| | Pt | N–top | −1.80 | 0.00 | −1.40 | 1.45 | N | 2.02 |
| | Au | N–top | −0.43 | 1.00 | −0.88 | 0.68 | N | 2.53 |

Systematic evaluation of adsorption along the groups of the periodic table reveals clear periodic trends. For Group 1 elements (H, Li, Na, K, Rb, **Figure 3**, left), adsorption on pristine h-BN is weak to moderate. Hydrogen binds at the bridge site with an adsorption energy of −0.05 eV, inducing a band gap of 4.13 eV and a magnetization of 1.0 $\mu_B$. Alkali metals stabilize at the bridge or B-top sites with nearly the same binding energies for Na, K, and Rb (−0.21 eV, only Rb is at B-top), and somewhat stronger Li binding (−0.31 eV). These atoms all maintain magnetization values of 1.0 $\mu_B$, consistent with their unpaired valence electrons. The band gap decreases systematically down the group, from 1.02 eV (Li) to 0.31 eV (Rb).

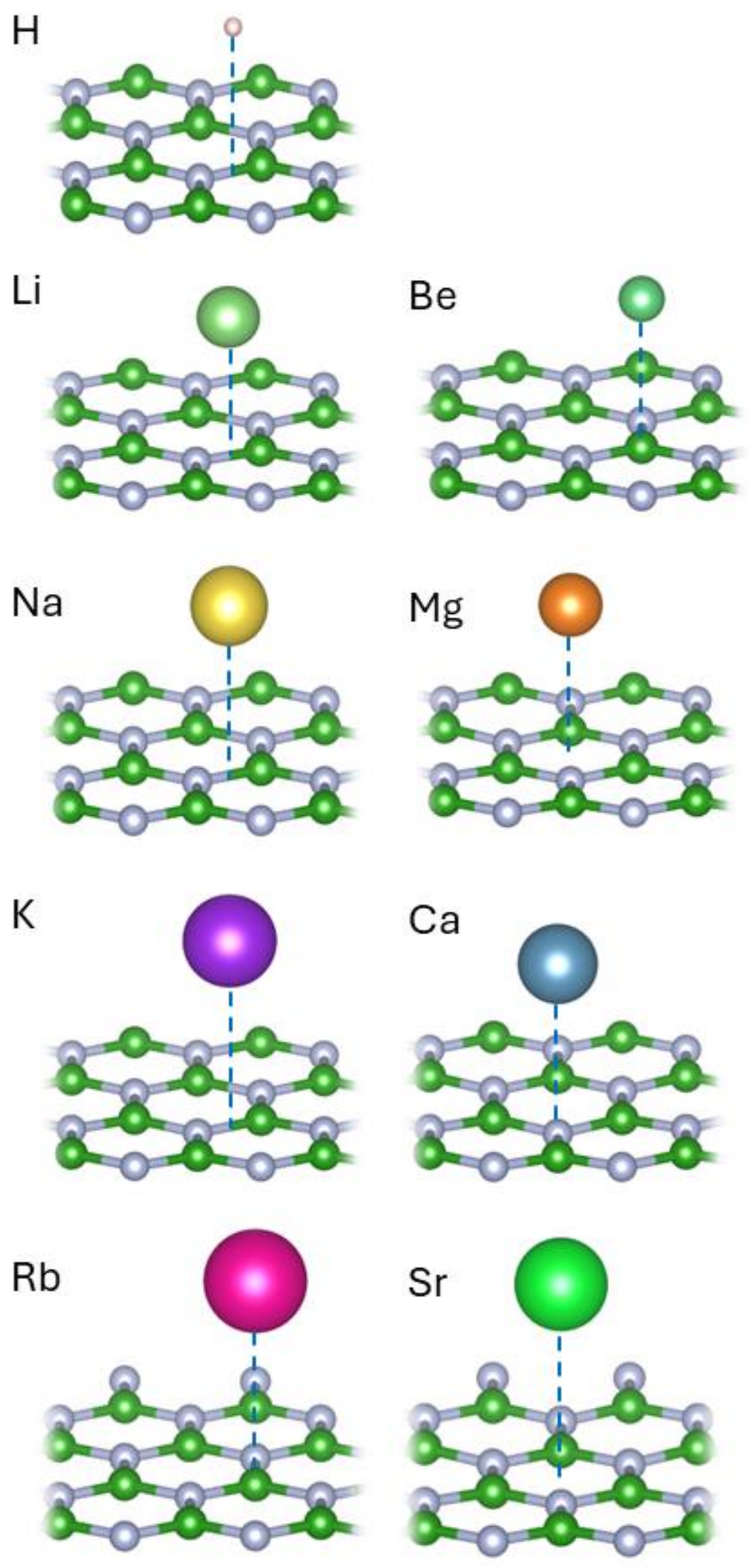

Figure 3. Optimized structures for hydrogen and alkaline metals (left column), and alkaline earth metals (right column) interacting with h-BN.

The Group 2 alkaline earth metals (Be, Mg, Ca, Sr, **Figure 3**, right column) show similar but, in most cases, slightly stronger interactions with h-BN. The adsorption energies range from −0.20 eV (Be, B-top) and −0.25 eV (Mg, hollow) to −0.31 eV (Ca, N-top), and −0.33 eV (Sr, hollow), showing a clear increase in interaction when moving down the group. All the systems are found to be nonmagnetic ($M$ = 0.0 $\mu_B$). Band gaps are 1.10-3.38 eV, generally showing a clear reduction with increasing atomic number.

Adsorption of group Ia and IIa elements on pristine h-BN lies firmly in the weak-interaction regime and is governed primarily by ionic size and adsorption distance rather than charge transfer or electronegativity. Despite modest electron donation, the large metal–surface separations suppress hybridization, resulting in negligible band-gap perturbation and uniformly weak binding. These results further emphasize the inert nature of defect-free h-BN and demonstrate that s-block elements cannot be effectively anchored without structural or chemical activation of the substrate.

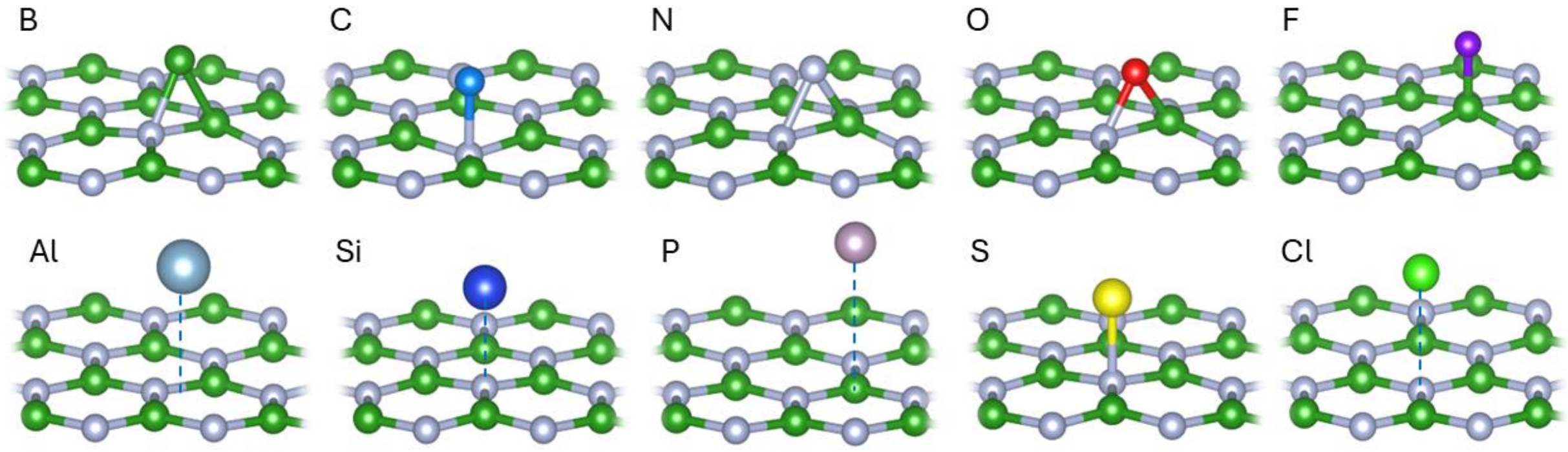

Figure 4. Optimized structures for p-elements in rows 2 and 3 of the PTE adsorbed onto h-BN.

Adsorption of p-block elements on pristine h-BN (**Figure 4**) exhibits clear periodic and chemical trends governed by the intrinsic polarity of the B–N bond and the electronic structure of the adsorbates. Second-row p-elements bind systematically stronger than their third-row counterparts, with the adsorption energies ranging from moderate chemisorption for O and F (≈ −2.0 eV) to weak binding for B and N (≈ −0.6 to −0.8 eV), whereas third-row elements generally interact weakly (≈ −0.1 to −1.1 eV). Electronegativity emerges as the primary descriptor: highly electronegative species (O, F,) preferentially adsorb at B-top or B-dominated bridge sites, act as electron acceptors, and induce substantial charge transfer ($|\Delta q| \geq 0.7$ e), resulting in short X–B bond lengths (≈1.4–1.5 Å) and pronounced band-gap narrowing due to the formation of adsorbate-derived states near the Fermi level. In contrast, more electropositive or open-shell adsorbates (C, Si) favor N-top sites, donate charge to the substrate, exhibit larger adsorption distances (>2 Å), and preserve significant magnetic moments, reflecting weak hybridization with the h-BN lattice. The magnitude of band-gap perturbation correlates with adsorption strength and charge transfer, being strongest for row-2 electronegative adsorbates and minimal for weakly bound row-3 elements. Thus, strong adsorption occurs only when significant charge transfer is accompanied by short bonding distance for smaller atoms, a condition satisfied predominantly by second-row electronegative elements. Larger third-row p-elements fail to meet these criteria simultaneously, resulting in weak binding and minimal electronic perturbation of the h-BN lattice.

As expected, noble gases (**Figure 5**) adsorb only very weakly onto h-BN. Helium at the hollow site binds with −0.02 eV and a band gap of 4.69 eV, while heavier noble gases bind more strongly. Ne has the binding energy of −0.04 eV (hollow, 4.67 eV band gap) while Ar deliberates −0.08 eV upon bonding at the hollow site (band gap 4.67 eV). All are nonmagnetic and leave the electronic structure essentially unperturbed.

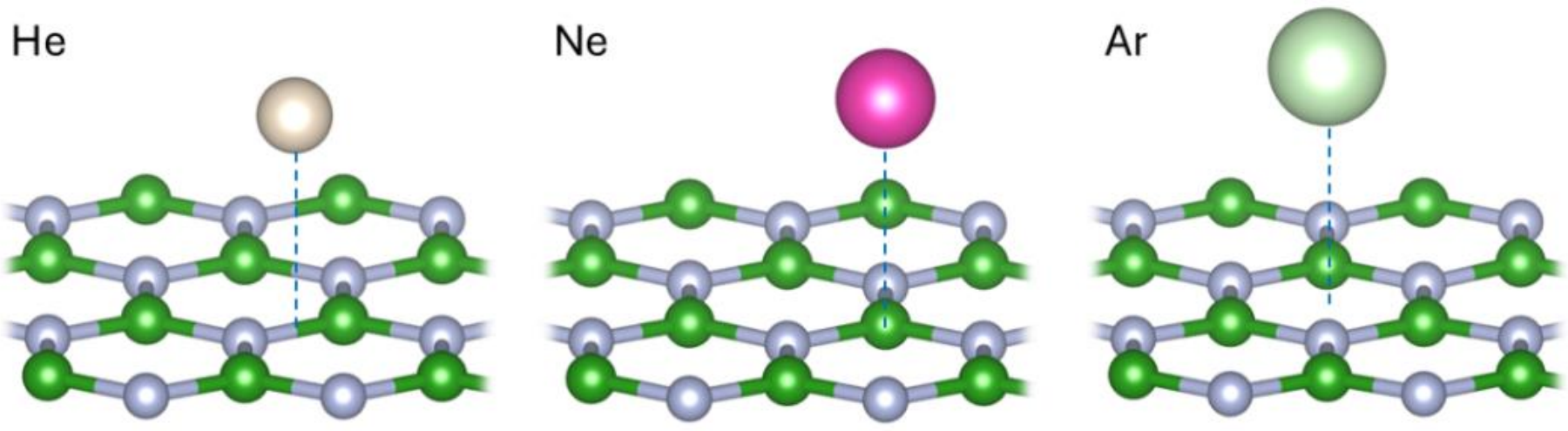

Figure 5. Optimized structures of noble gases interacting with h-BN.

In contrast to p-block adsorbates, d-metal adsorption on pristine h-BN (**Figure 6**) is governed primarily by d–p hybridization rather than charge transfer. Transition metals with partially filled d-bands (Ni, Ru, Rh, Ir, Pt) exhibit moderate to strong chemisorption, with adsorption energies exceeding −1.3 eV despite relatively large metal–substrate distances (≈2.0–2.3 Å). The absence of a monotonic correlation between adsorption energy and charge transfer highlights the secondary role of $\Delta q$ for d-metals, which instead reflects redistribution associated with bond formation. Coinage metals (Cu, Ag, Au), characterized by filled $d^{10}$ shells, interact weakly with h-BN, retain adsorption distances > 2Å, and induce minimal electronic perturbation. These findings demonstrate a fundamental change in adsorption descriptors when moving from p-elements to d-metals, emphasizing d-band-driven hybridization as the key factor controlling metal anchoring on defect-free h-BN.

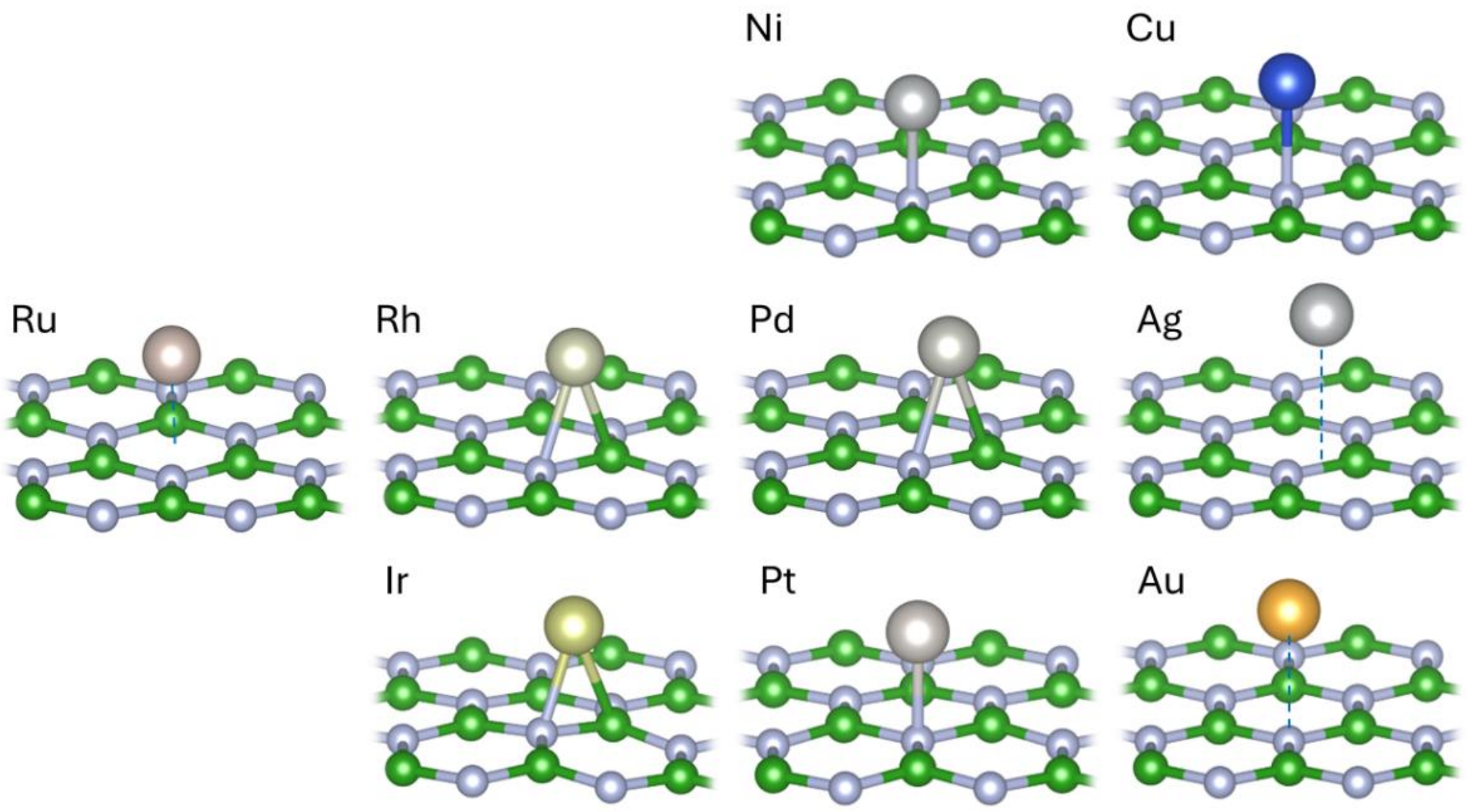

Figure 6. Optimized structures of selected d-elements and coinage metals adsorbed onto h-BN.

To conclude so far, adsorption on pristine h-BN is weak for most elements, with only a limited subset, most notably oxygen, fluorine, and carbon among the p-block, showing strong chemisorption. Across individual groups, adsorption energies generally decrease with increasing atomic size, reflecting reduced orbital overlap and increased adsorption distances. Band gaps are reduced most significantly by highly electronegative adsorbates and by some of the late transition metals. Magnetization is preserved for open-shell atoms, including alkali metals, Group 15 elements, and some transition metals, while closed-shell species remain nonmagnetic. **Figure 7** presents DOSs for representative atoms adsorbed on pristine h-BN, illustrating the characteristic range from physisorption to strong chemisorption. The DOS of the clean surface shows the wide intrinsic band gap of h-BN (**Figure 1**), confirming its electronic inertness. Adsorption of a weakly interacting species such as Ar leaves the electronic structure essentially unchanged, reflecting purely van der Waals binding. For Ni, moderate adsorption leads to the appearance of narrow d-derived states near the Fermi level, slightly reducing the band gap and indicating limited hybridization between Ni 3d and B/N 2p orbitals. In contrast, the strongly bound F adatom introduces pronounced localized states within the gap and shifts the band edges, consistent with the large adsorption energy and significant modification of the electronic properties.

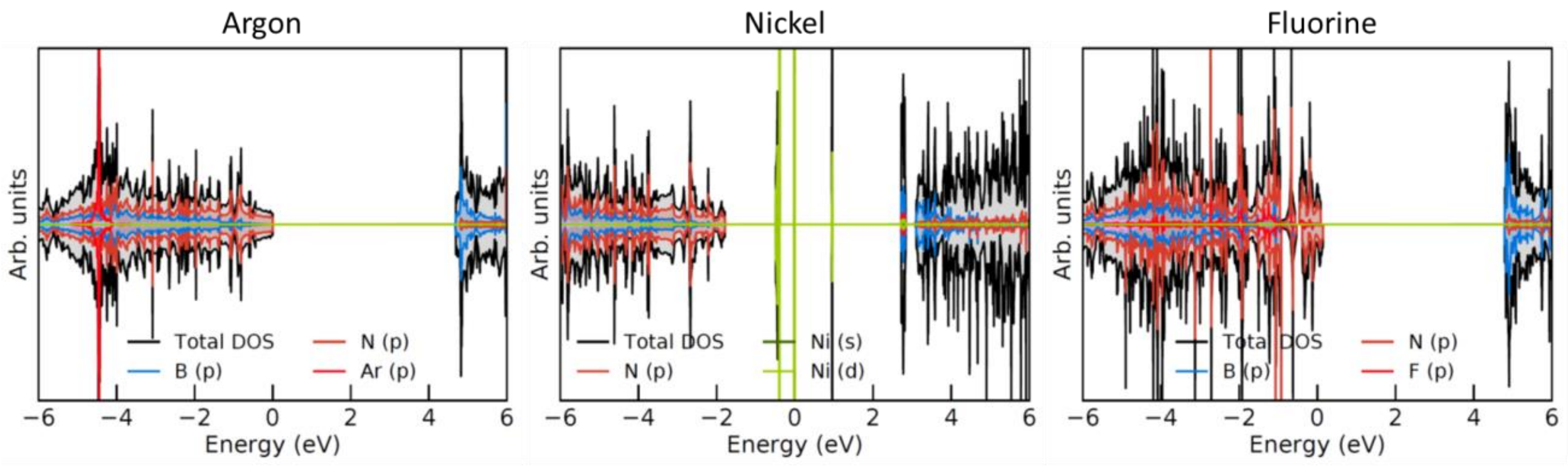

Figure 7. Representative DOS plots for atomic adsorption on pristine h-BN: weakly bound Ar, intermediately strong bonded Ni, and strongly bound fluorine.

### 3.3 Adsorption at the boron vacancy ($V_B$)

The calculated binding energies for adsorption at the $V_B$ site are summarized in **Table 2**. The introduction of a boron vacancy dramatically enhances the chemical reactivity of h-BN, as evidenced by adsorption energies that are significantly larger than those obtained on the pristine surface. In nearly all cases, binding energies exceed −1 eV and, for many elements, surpass the cohesive energy of the corresponding elemental phase, indicating strong chemisorption and the thermodynamic stabilization of single atoms or substitution-like configurations at the vacancy. This behavior reflects a fundamental change in the adsorption mechanism, from weak polarization-driven interactions on pristine h-BN to covalent bond formation with the three undercoordinated nitrogen atoms surrounding the vacancy.

Table 2. Binding energies ($E_b$) of element X onto B-vacancy in h-BN, corresponding total magnetizations of the obtained doped h-BN ($M_{tot}$), charge transferred from X to the surrounding atoms ($\Delta q$(X)), band gaps ($E_{gap}$), and the distances $d$ between X and vacancy's first N neighbor. **Bold values** of binding energies show that the $|E_b(X)|$ > cohesive energy of element X.

| PTE part | X | $E_b$(X) / eV | $M_{tot}$ / $\mu_B$ | $\Delta q$(X) / e | $E_{gap}$ / eV | $d$(X–N) / Å |
|---|---|---|---|---|---|---|
| group Ia | H | −4.94 | 2.00 | 0.45 | 0.15 | 1.03 |
| | Li | **−5.23** | 2.00 | 0.88 | 0.91 | 1.94 |
| | Na | **−4.05** | −2.00 | 0.56 | 1.02 | 2.31 |
| | K | **−4.04** | 2.00 | 0.86 | 1.17 | 2.67 |
| | Rb | **−3.95** | 2.00 | 0.87 | 1.18 | 2.82 |
| group IIa | Be | **−10.49** | 0.19 | 1.65 | 0.02 | 1.55 |
| | Mg | **−5.84** | 1.00 | 1.59 | 0.00 | 1.93 |
| | Ca | **−7.01** | 1.00 | 1.36 | 0.16 | 2.23 |
| | Sr | **−6.50** | 1.00 | 1.45 | 0.10 | 2.37 |
| p – row 2 | B | **−16.45** | 0.00 | 2.11 | 4.70 | 1.45 |
| | C | **−13.76** | 1.00 | 1.09 | 0.89 | 1.42 |
| | N | **−8.73** | 0.00 | 0.10 | 3.38 | 1.49 |
| | O | **−5.76** | 1.00 | −0.21 | 1.17 | 1.47 |
| | F | **−1.74** | 0.00 | −0.14 | 1.90 | 1.71 |
| p – row 3 | Al | **−10.24** | 0.00 | 2.37 | 3.91 | 1.70 |
| | Si | **−11.71** | 1.00 | 2.30 | 1.46 | 1.71 |
| | P | **−10.58** | 0.00 | 2.02 | 3.73 | 1.72 |
| | S | **−6.87** | 1.00 | 1.11 | 0.93 | 1.68 |
| | Cl | **−2.43** | 0.00 | 0.06 | 0.11 | 1.72 |
| noble | He | −0.07 | 1.00 | −0.01 | 0.40 | 3.07 |
| | Ne | **−0.08** | 1.00 | −0.01 | 0.41 | 3.17 |
| | Ar | **−0.11** | 1.00 | −0.01 | 0.78 | 3.45 |
| d – row 4 | Ni | **−8.37** | 1.00 | 0.90 | 0.00 | 1.79 |
| | Cu | **−6.06** | 2.00 | 0.84 | 0.75 | 1.83 |
| d – row 5 | Ru | **−10.21** | 1.00 | 0.82 | 0.37 | 1.91 |
| | Rh | **−9.26** | 0.00 | 0.79 | 1.68 | 1.90 |
| | Pd | **−5.77** | 1.00 | 0.59 | 0.01 | 1.96 |
| | Ag | **−3.66** | 2.00 | 0.60 | 0.83 | 2.10 |
| d – row 6 | Ir | **−9.48** | 0.00 | 0.55 | 1.71 | 1.92 |
| | Pt | **−7.11** | 1.00 | 0.59 | 0.00 | 1.97 |
| | Au | −3.79 | 2.00 | 0.47 | 0.73 | 2.05 |

For Group 1 elements (H, Li, Na, K, Rb, **Figure 8**), adsorption energies range from −5.23 eV for lithium to approximately −4.0 eV for the heavier alkali metals, representing an increase of nearly an order of magnitude compared to pristine h-BN. Despite their low electronegativity, these elements bind strongly by donating charge to the vacancy-induced N dangling bonds, as reflected in substantial charge transfer ($\Delta q \approx 0.5$–$0.9$ e) and short X–N distances. The total magnetization remains finite ($\approx 2\ \mu_B$), consistent with partial occupation of defect-related states. The band gap is reduced to around 1 eV for Li, Na, K, and Rb, indicating incomplete passivation of the vacancy and the persistence of localized in-gap states. Group 2 elements (Be, Mg, Ca, Sr, **Figure 8**) exhibit even stronger binding, with adsorption energies ranging from −10.5 eV for Be to −5.8 eV for Mg. These values exceed the cohesive energies of the bulk metals [29], demonstrating that incorporation into the vacancy is energetically favored over metal aggregation. The interaction is characterized by large charge transfer ($\Delta q \approx 1.3$–$1.7$ e) and short X–N distances, confirming strong covalent bonding. Magnetization is low compared to non-saturated $V_B$ (0–1 $\mu_B$), with band gaps collapsing to near-metallic values in some cases (e.g., 0.02 eV for Be), highlighting the efficiency of vacancy states in coupling to s-block adsorbates.

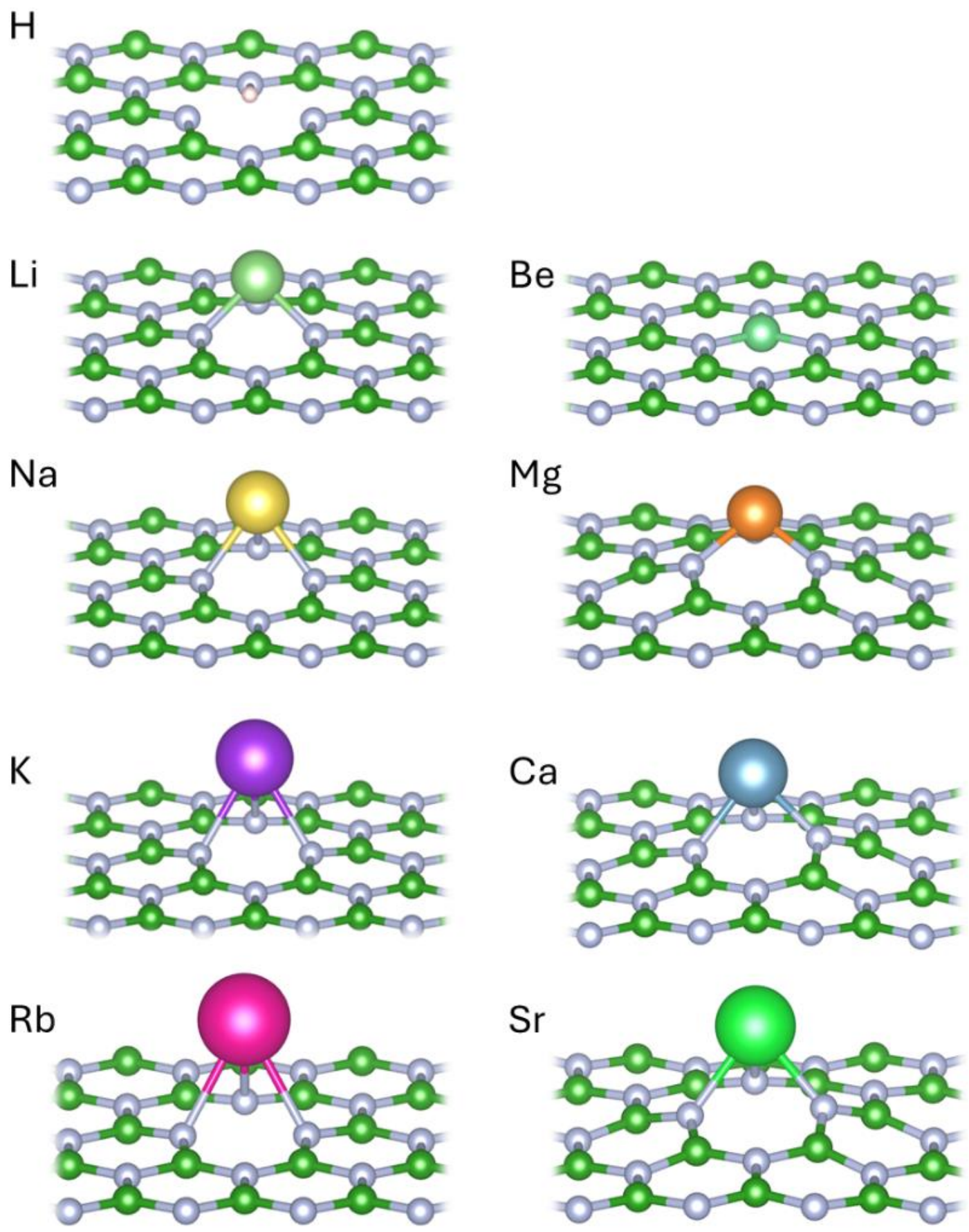

Figure 8. Optimized structures for hydrogen, alkaline, and alkaline earth metals interacting with B-vacancy in h-BN.

Group 13 elements also display very large binding energies, from −10.2 eV for Al to −16.4 eV for B. These atoms effectively saturate the dangling bonds of the vacancy, leading to closed-shell, nonmagnetic configurations and a recovery of wide band gaps (≈3.9–4.7 eV). This behavior indicates that Group 13 adsorption can electronically "heal" the boron vacancy, restoring semiconducting character to the h-BN lattice. Strong adsorption persists for Group 14 elements. Carbon binds with −13.76 eV and produces a moderate band gap of 0.89 eV, while silicon adsorbs with −11.71 eV and a gap of 1.46 eV. The slight reduction in adsorption strength down the group correlates primarily with increasing atomic size and longer X–N bond lengths, rather than changes in electronegativity, emphasizing that geometric compatibility with the vacancy plays a key role. Group 15 elements (N, P) also exhibit very strong binding (−8.73 eV for N and −10.58 eV for P), accompanied by zero magnetization and relatively large band gaps (≈3–4 eV), indicative of effective vacancy passivation. For chalcogens (Group 16), oxygen and sulfur bind strongly (−5.76 and −6.87 eV, respectively), but with smaller band gaps (≈0.9–1.2 eV), reflecting partial filling of defect-derived states. In Group 17, fluorine and chlorine bind less strongly than oxygen and sulfur but still well within the chemisorption regime (−1.74 and −2.43 eV), producing small band gaps and nonmagnetic ground states. Optimized structures are shown in **Figure 9**.

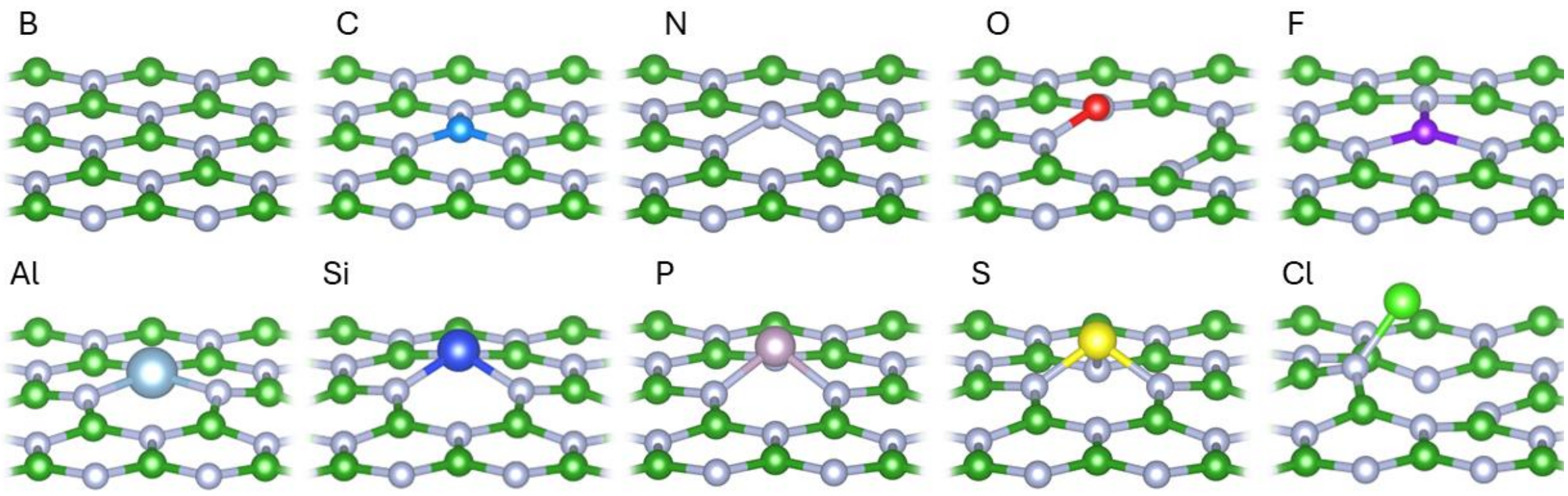

Figure 9. Optimized structures for row 2 and 3 p-elements of the PTE adsorbed onto B-vacancy in h-BN.

The noble gases (He, Ne, Ar, **Figure 10**) display the weakest interaction with the vacancy. Adsorption energies are from −0.07 eV to −0.11 eV, indicating purely dispersion-dominated binding. Their band gaps are in the range 0.4–0.8 eV, and magnetizations are 1 $\mu_B$ for all the systems.

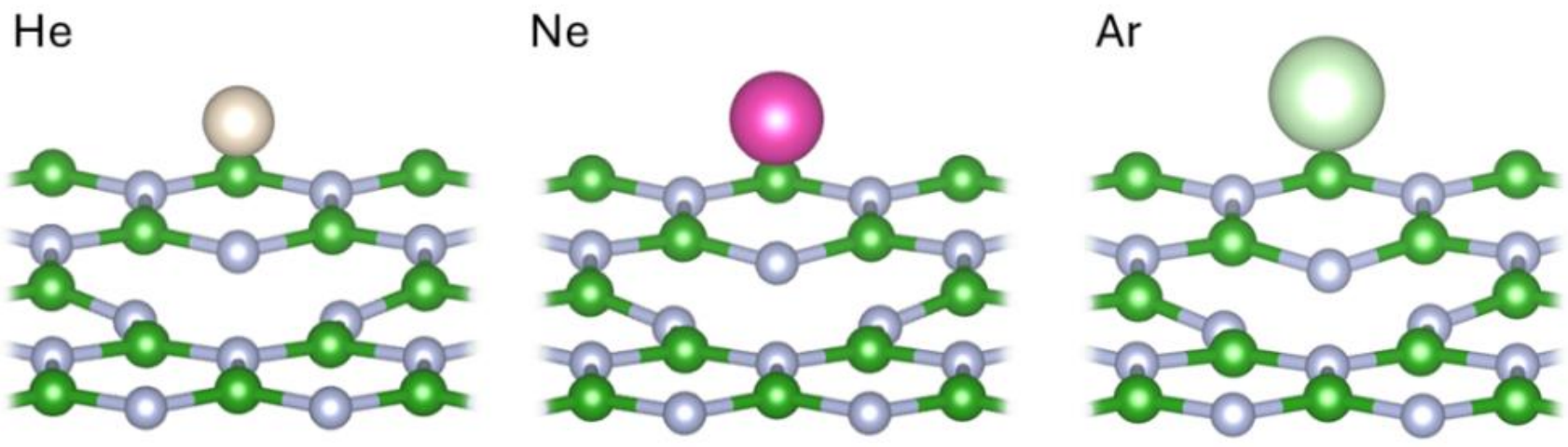

Figure 10. Optimized structures of noble gases interacting with B-vacancy in h-BN.

Transition and coinage metals (**Figure 11**) show a much wider and significantly stronger adsorption range. Late transition metals (Ni, Ru, Rh, Pd, Ir, Pt) bind very strongly, with adsorption energies around −8 to −10 eV, while coinage metals (Cu, Ag, Au) have somewhat weaker but still notable adsorption (−6 to −3.8 eV). These values are much higher (comparing absolute values) than on pristine h-BN, emphasizing the key role of the boron vacancy in stabilizing metallic species through strong interactions with undercoordinated nitrogen atoms. Unlike on the pristine surface, the adsorption strength of d-metals at the vacancy does not increase steadily with charge transfer. Instead, it depends on the level of d–p hybridization and how well the metal can saturate the vacancy-related dangling bonds. The calculated total magnetizations range from 0 to 2 $\mu_B$, showing different levels of hybridization between metal d states and vacancy-induced states. Highly hybridized systems (e.g., Ni, Pd, Pt) tend to have quenched or reduced magnetic moments, while weaker coupling allows some atomic magnetism to remain. The electronic structures often become nearly metallic, with very small or vanishing band gaps (<0.01 eV for Ni, Pd, and Pt, around 0.7 eV for Ag and Au), caused by metal-related states at the Fermi level.

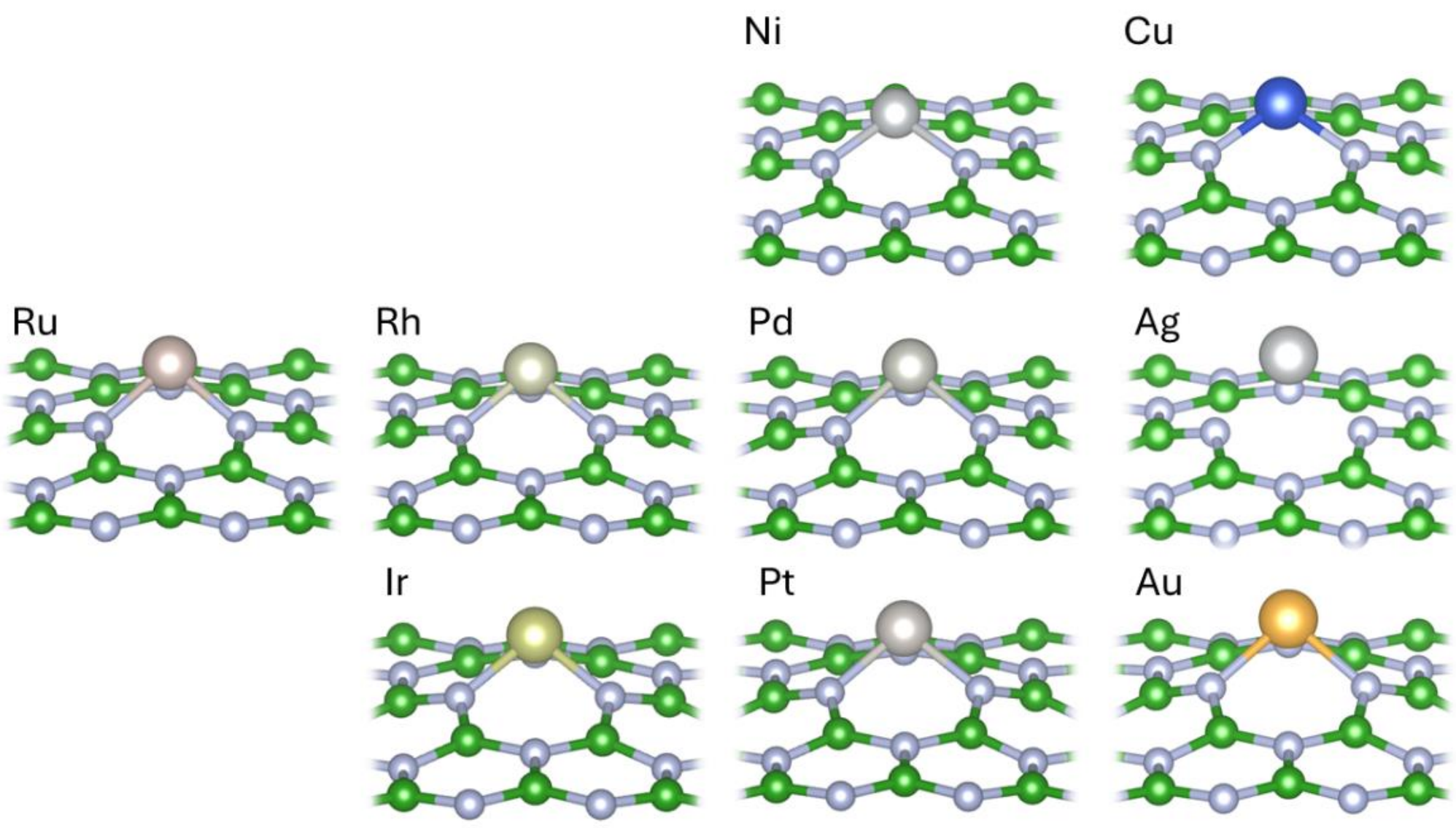

Figure 11. Optimized structures of selected d-elements and coinage metals adsorbed onto B-vacancy of h-BN.

The adsorption at the boron vacancy is strongly exothermic for nearly all elements except the noble gases. The energy range covers almost two orders of magnitude, from approximately −0.1 eV for physisorption of noble gases (He, Ne, Ar) to about −16 eV for the elements with the strongest binding (B, C). Band gaps are notably reduced compared to pristine h-BN, and metallic or near-metallic behavior often occurs with transition-metal and light-element adsorption. Magnetization generally decreases after adsorption, with localized moments only remaining for open-shell atoms. These findings show that the boron vacancy turns the normally inert h-BN lattice into a highly reactive site capable of stabilizing a wide variety of species.

### 3.4 Adsorption at the nitrogen vacancy ($V_N$)

The calculated binding energies for adsorption at the nitrogen vacancy site are summarized in **Table 3**. The $V_N$ site creates a local environment consisting of three undercoordinated boron atoms, which is fundamentally different from the nitrogen-rich coordination present at the $V_B$ site. As a consequence, adsorption at $V_N$ is governed primarily by interactions between the adsorbate and electron-deficient boron atoms, resulting in stabilization patterns that are complementary to those observed at $V_B$. Overall, adsorption energies at $V_N$ are systematically lower than at $V_B$, confirming that nitrogen vacancies are moderately reactive but more selective adsorption sites.

Only a limited subset of elements binds at $V_N$ with energies comparable to or exceeding their cohesive energies, most notably light p-block elements and selected late transition metals. In contrast, alkali and alkaline-earth metals are only moderately stabilized and frequently relax away from the vacancy center. This behavior reflects the lower electronegativity and weaker orbital overlap of the boron atoms surrounding $V_N$, which limits the strength of covalent bonding compared to the nitrogen-terminated environment of $V_B$.

Table 3. Binding energies ($E_b$) of element X onto N-vacancy in h-BN, corresponding total magnetizations ($M_{tot}$) of the obtained doped h-BN, charge transferred from X to the surrounding atoms ($\Delta q$(X)), band gaps ($E_{gap}$), and the distances $d$ between X and vacancy's first N neighbor. **Bold values** of binding energies show that the $|E_b(X)|$ > cohesive energy of element X.

| PTE part | X | $E_b$(X) / eV | $M_{tot}$ / $\mu_B$ | $\Delta q$(X) / e | $E_{gap}$ / eV | $d$ / Å |
|---|---|---|---|---|---|---|
| group Ia | H | −3.96 | 0.00 | −1.01 | 3.57 | 1.21 |
| | Li | −1.51 | 0.00 | 0.88 | 1.57 | 2.35 |
| | Na | −1.01 | 0.00 | 0.72 | 1.23 | 2.62 |
| | K | −0.91 | 0.00 | 0.76 | 0.65 | 2.98 |
| | Rb | **−0.91** | 0.00 | 0.71 | 0.64 | 3.07 |
| group IIa | Be | −2.08 | 1.00 | 1.41 | 0.97 | 1.77 |
| | Mg | −0.71 | 1.00 | 0.42 | 0.74 | 2.50 |
| | Ca | −1.12 | 1.00 | 0.32 | 0.53 | 2.81 |
| | Sr | −1.03 | 1.00 | 0.57 | 0.47 | 3.00 |
| p – row 2 | B | **−7.14** | 0.00 | 0.45 | 1.19 | 1.62 |
| | C | **−11.42** | 1.00 | −2.04 | 0.74 | 1.51 |
| | N | **−12.83** | 0.00 | −2.18 | 4.69 | 1.45 |
| | O | **−9.55** | 1.00 | −1.61 | 1.00 | 1.46 |
| | F | **−6.61** | 0.00 | −0.85 | 3.75 | 1.37 |
| p – row 3 | Al | −2.03 | 0.00 | 0.66 | 0.97 | 2.29 |
| | Si | **−4.98** | 1.00 | 1.03 | 0.63 | 1.99 |
| | P | **−7.30** | 0.00 | −0.43 | 3.99 | 1.88 |
| | S | **−6.32** | 1.00 | −1.32 | 0.54 | 1.91 |
| | Cl | **−4.67** | 0.00 | −0.94 | 3.77 | 1.83 |
| noble | He | −0.06 | 1.00 | −0.01 | 0.84 | 2.98 |
| | Ne | **−0.08** | 1.00 | −0.02 | 0.77 | 3.30 |
| | Ar | **−0.11** | 1.00 | −0.01 | 0.78 | 3.50 |
| d – row 4 | Ni | **−5.64** | 1.00 | −0.48 | 0.61 | 1.90 |
| | Cu | −2.87 | 0.00 | −0.28 | 1.69 | 2.18 |
| d – row 5 | Ru | **−7.64** | 1.00 | −0.57 | 0.40 | 2.01 |
| | Rh | **−8.18** | 0.00 | −0.68 | 2.65 | 1.96 |
| | Pd | **−4.96** | 1.00 | −0.70 | 0.56 | 2.03 |
| | Ag | −2.144 | 0.00 | −0.21 | 1.85 | 2.41 |
| d – row 6 | Ir | **−9.14** | 0.00 | −2.37 | 2.47 | 1.97 |
| | Pt | **−7.36** | 1.00 | −1.89 | 0.76 | 2.02 |
| | Au | −3.67 | 0.00 | −0.51 | 2.30 | 2.26 |

The s-block elements (Groups 1 and 2, **Figure 12**) interact relatively weakly with $V_N$. Their adsorption energies ($|E_b|$ ≈ 1–2 eV) are several times smaller than those at $V_B$, and total magnetization typically remains quenched or small, consistent with limited orbital hybridization. The corresponding band gaps remain relatively wide (≈ 0.5–1.5 eV), but they narrow as one goes down the groups.

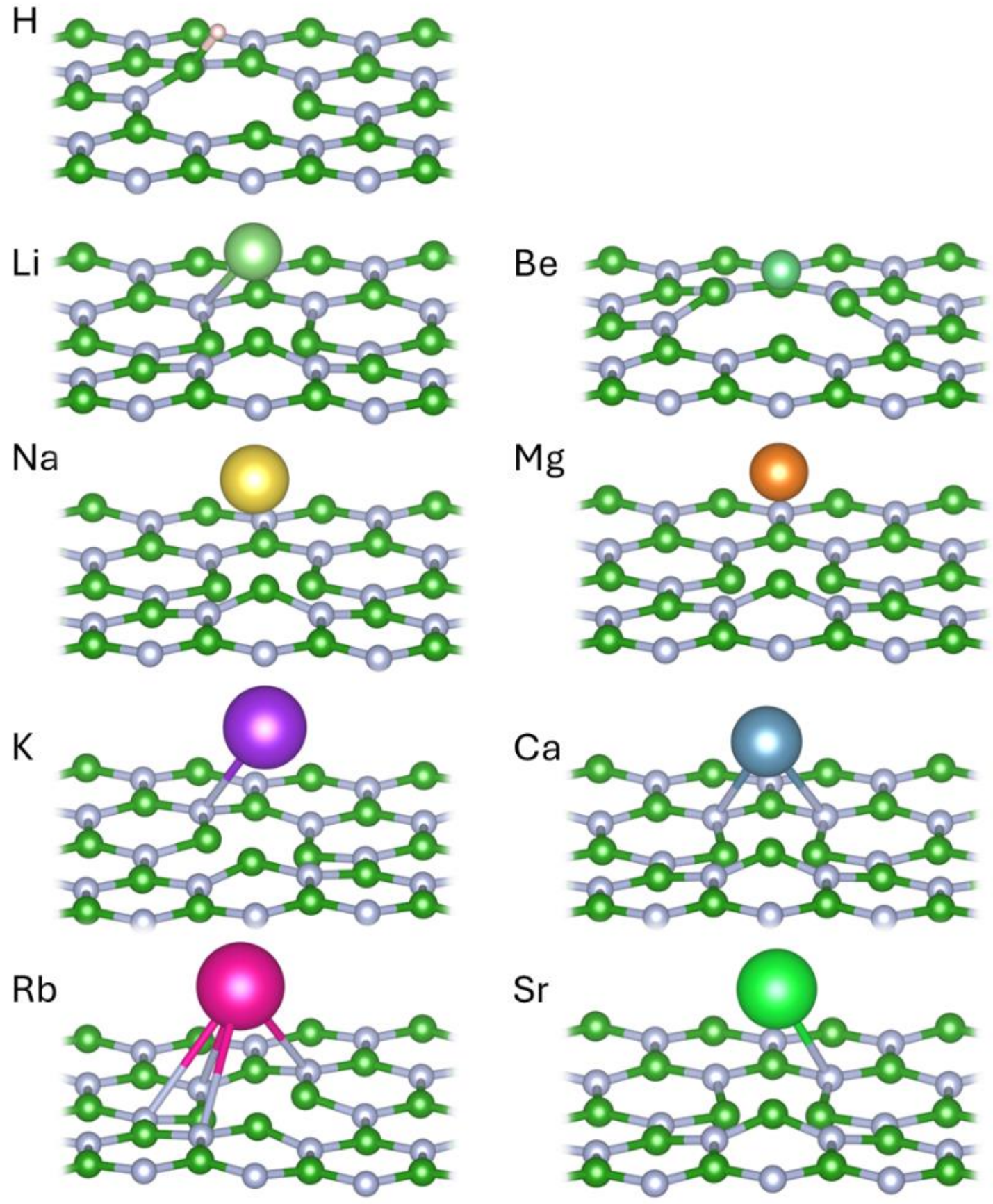

Figure 12. Optimized structures for hydrogen, alkaline, and alkaline earth elements interacting with N-vacancy in h-BN.

Among the p-block elements (**Figure 13**), the second-row species (C, N, O, F) exhibit the strongest adsorption. Carbon and nitrogen form the most stable vacancy complexes, with adsorption energies exceeding −10 eV, indicative of pronounced local reconstruction and strong directional B–X bonding. Oxygen also binds very strongly (−9.55 eV), while fluorine exhibits slightly weaker but still substantial stabilization (−6.61 eV). In contrast, heavier congeners in the same groups (S, Cl) display reduced adsorption energies, consistent with increasing atomic size and diminished orbital overlap with the boron dangling bonds. In all cases of strong chemisorption, adsorption leads to pronounced band-gap narrowing, frequently below 1 eV, reflecting the introduction of occupied and unoccupied defect-derived states near the Fermi level.

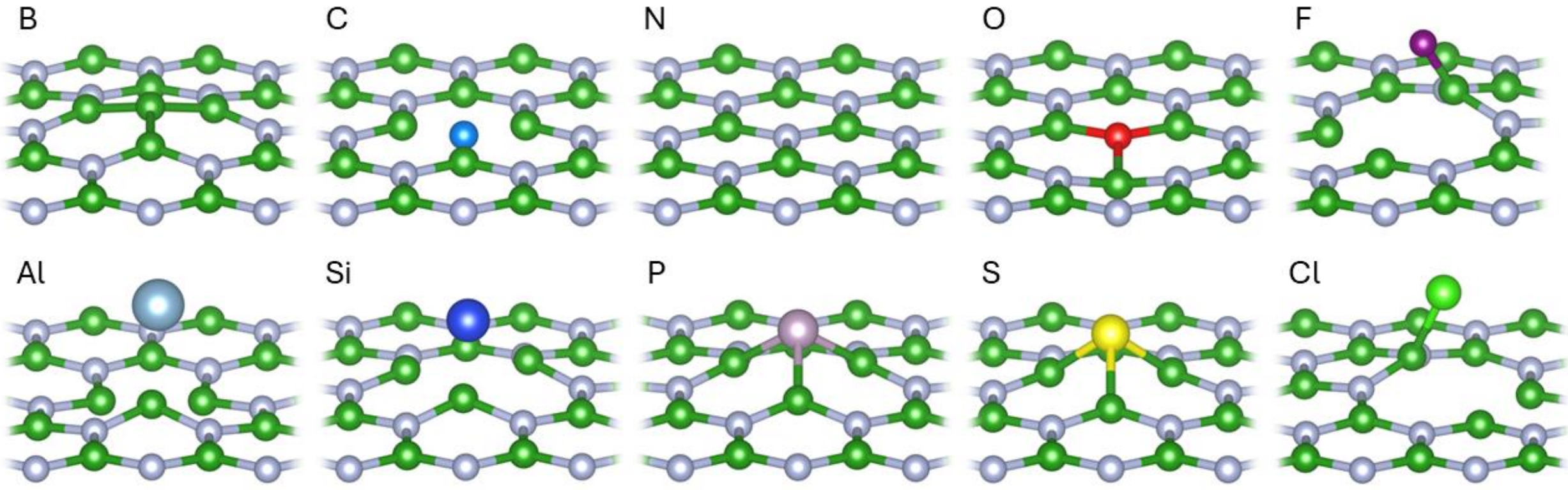

Figure 13. Optimized structures row 2 and 3 p-elements of the PTE adsorbed onto N-vacancy in h-BN.

Noble gases (**Figure 14**) interact only weakly with $V_N$, with adsorption energies below −0.1 eV and large adsorption distances, confirming purely dispersion-dominated physisorption.

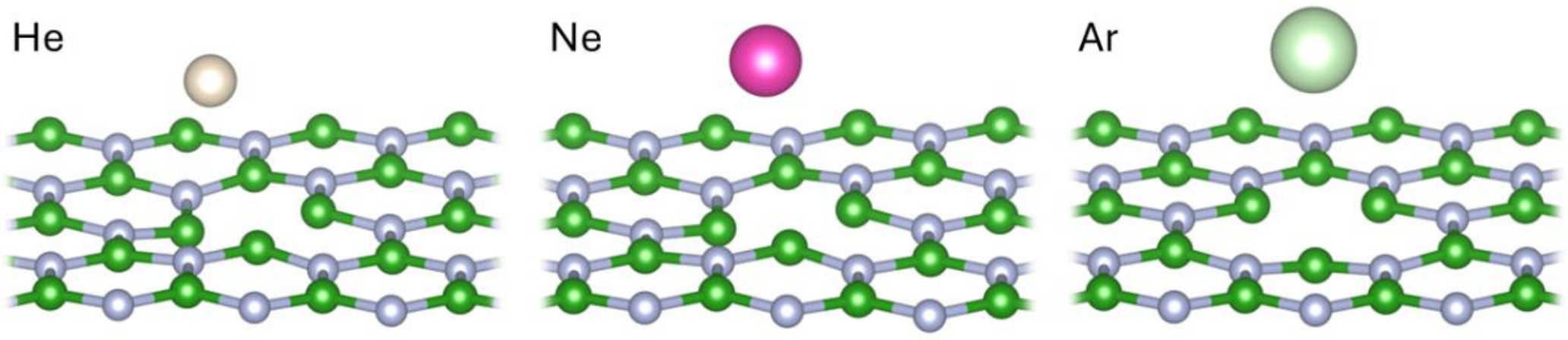

Figure 14. Optimized structures of noble gases interacting with N-vacancy in h-BN.

Transition-metal adsorption at $V_N$ (**Figure 15**) remains strongly exothermic but is generally weaker than at the boron vacancy. The largest adsorption energies are obtained for Ru, Rh, Ir, and Pt ($|E_b| \approx$ 7–9 eV), reflecting efficient coordination to the three undercoordinated boron atoms and substantial d–p hybridization. Coinage metals (Cu, Ag, Au) bind weaker ($|E_b| \approx$ 2–4 eV), and their larger band gaps (≈ 1.5–2.3 eV) indicate less effective coupling to the vacancy. Magnetization is generally quenched for late transition metals, while finite moments persist for selected open-shell configurations, reflecting incomplete hybridization.

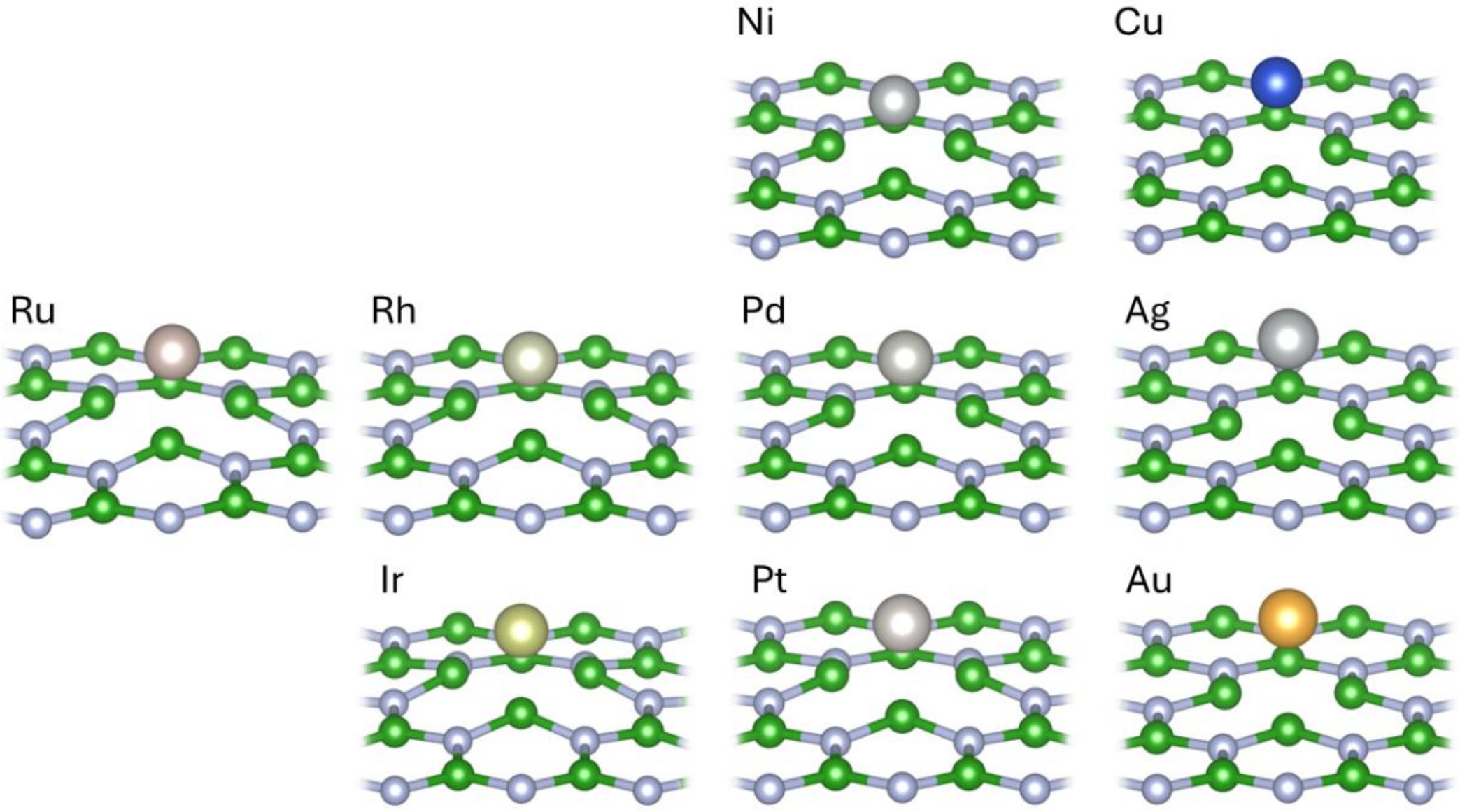

Figure 15. Optimized structures of selected d-elements and coinage metals adsorbed onto N-vacancy of h-BN.

The electronic structures of representative adsorbates on defective h-BN surfaces are summarized in **Figure 16**. The DOS clearly shows that vacancy formation profoundly modifies the electronic structure relative to pristine h-BN by introducing defect-derived states within the wide intrinsic band gap. Adsorption further reshapes these states, with the extent depending strongly on both the vacancy type and the chemical nature of the adsorbate. For argon, the DOS remains essentially unchanged for both boron and nitrogen vacancies, with Ar-derived states located far from the Fermi level and negligible hybridization with the substrate, confirming purely dispersion-dominated physisorption. In contrast, adsorption of Ni at the boron vacancy leads to strong hybridization between Ni 3d and N 2p states, producing broad, delocalized features that cross the Fermi level and yield a nearly metallic electronic structure, consistent with the vanishing band gap obtained from total-energy calculations. At the nitrogen vacancy, Ni-derived states remain close to the Fermi level but are more localized and asymmetric, resulting in pronounced gap narrowing rather than full metallization, reflecting weaker d–p coupling with the surrounding boron atoms. Adsorption of fluorine exhibits a qualitatively different behavior. At the boron vacancy, F introduces sharp, localized p-derived states within the band gap, significantly perturbing the electronic structure while preserving a finite gap and semiconducting character. At the nitrogen vacancy, these defect states become more localized and shift further from the Fermi level, consistent with reduced covalent stabilization due to the electron-deficient boron coordination.

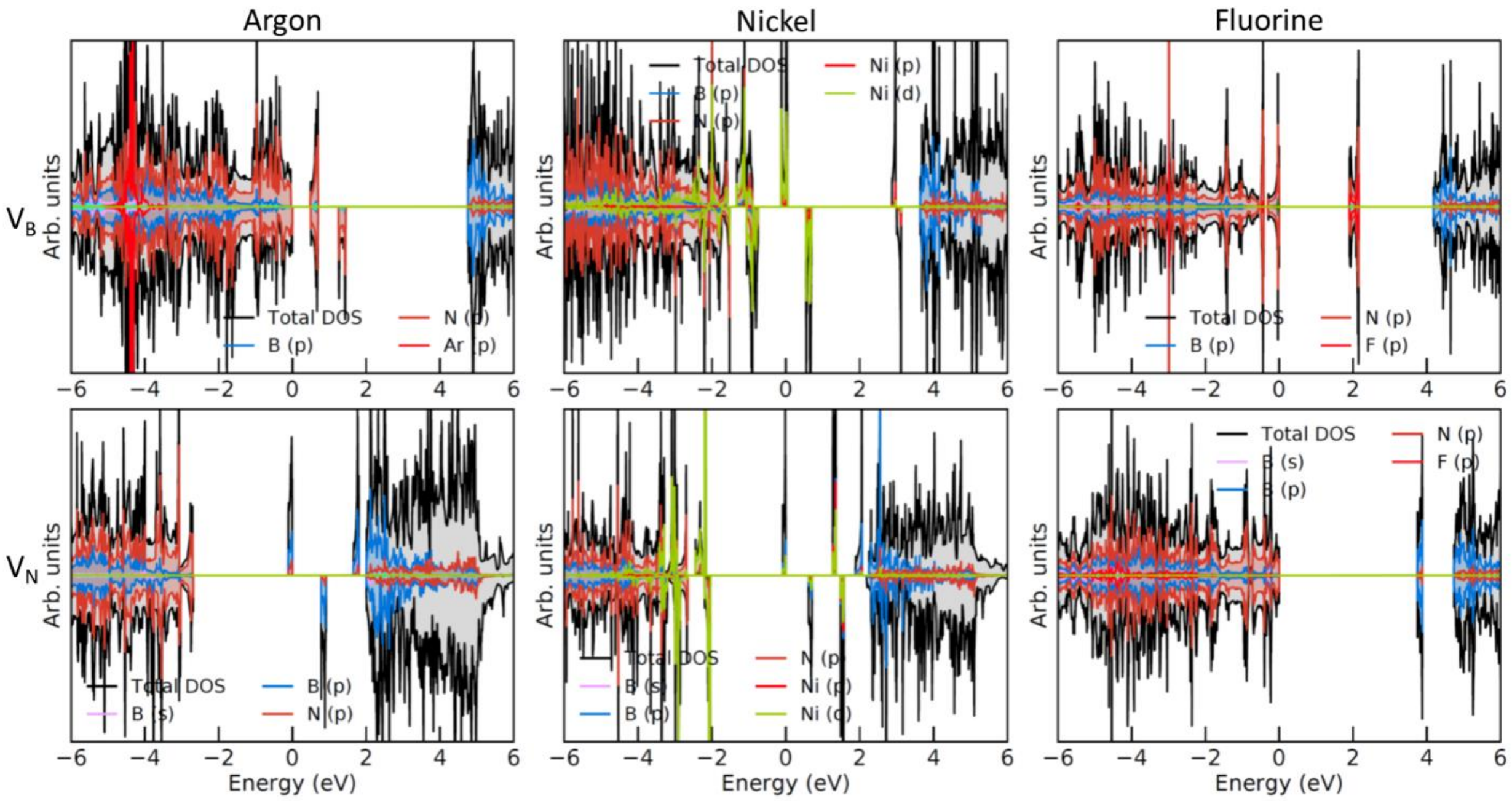

Figure 16. DOS plots for Ar, Ni, and F binding to boron vacancy (top row) and nitrogen vacancy (bottom row), in the h-BN lattice.

### 3.5 Periodic trends across the PTE

Analysis of adsorption across the periodic table reveals systematic and chemically consistent trends for pristine and defected h-BN. On the pristine surface, the closed electronic structure of the B–N network results in minimal intrinsic reactivity: only light, highly electronegative elements such as O and F, as well as small covalently bonding species such as C, exhibit significant adsorption energies, while most elements interact predominantly through weak dispersion forces. Consequently, the band gap of pristine h-BN remains wide (≈4–5 eV), except in these limited cases where localized states appear near the band edges (**Figure S2**, Supplementary Information).

The introduction of vacancies fundamentally alters this picture. At the $V_B$ site, the three undercoordinated nitrogen atoms act as centers that strongly stabilize electropositive and metallic adsorbates. As a result, adsorption energies are maximized for s-block elements and transition metals, frequently exceeding the cohesive energies of the corresponding elemental phases. Late transition and coinage metals (Ni, Pd, Pt, Cu, Ag, Au) form particularly stable configurations, often approaching metallic electronic structures, while light p-block elements such as C and O also bind exceptionally strongly. In most cases, adsorption at $V_B$ leads to substantial band-gap reduction (**Figure S3**, Supplementary Information) and, for many transition-metal systems, to the emergence of a finite density of states at the Fermi level. Magnetization is generally quenched, reflecting strong hybridization between the adatom states and nitrogen lone-pair orbitals.

In contrast, the nitrogen vacancy $V_N$ is surrounded by three electron-rich boron atoms and therefore exhibits a complementary selectivity. Strong chemisorption is observed primarily for electronegative and covalently bonded adsorbates such as C, N, O, and F, whereas metallic species and alkali or alkaline-earth elements are less stabilized. Transition-metal adsorption at $V_N$ remains significant but is systematically weaker than at $V_B$, consistent with the reduced availability of electron density on boron. The electronic structures of $V_N$ systems typically retain finite band gaps (**Figure S4**, Supplementary Information), and discrete mid-gap states persist, indicating more localized electronic reconstruction than at the boron vacancy. Finite magnetic moments are often preserved for open-shell adsorbates such as Ru, Pd, and Pt, consistent with their weaker hybridization.

Across the periodic table, in most cases, the adsorption strength decreases down each group, reflecting increasing atomic radius and diminishing orbital overlap with the substrate. This trend is most pronounced for p-block elements, where second-row members (C, N, O, F) form strong, localized bonds at defect sites, while heavier congeners interact more weakly. The systematic comparison of pristine, $V_B$, and $V_N$ surfaces thus demonstrates that defect type governs both the magnitude and the nature of adsorption: $V_B$ acts as a strongly electron-donating site that favors metallic and electropositive species, whereas $V_N$ behaves as a more selective center that stabilizes electronegative atoms and directional covalent bonding motifs.

To provide a quantitative measure of adsorption stability at the vacancy sites, **Figure 17** compares the calculated binding energies of each element at the boron and nitrogen vacancies with their respective cohesive energies in the condensed phase. This comparison captures the balance between adatom-substrate and atom-atom interactions and outlines the boundary between strong chemisorption ($|E_b|>E_{coh}$) and weak binding or physisorption ($|E_b|<E_{coh}$). For the boron vacancy, most elements lie well above the diagonal line defined by $|E_b|=E_{coh}$, indicating that interaction with the defected surface is energetically favored over clustering in the bulk. This effect is especially pronounced for light p-block elements (B, C, N, O) and late transition metals (Ni, Ru, Rh, Pd, Ir, Pt), whose binding energies reach 8–16 eV and substantially exceed their cohesive energies. These species, therefore, exhibit a strong thermodynamic driving force to remain immobilized at the vacancy, suggesting that isolated adatoms can be stabilized under equilibrium conditions.

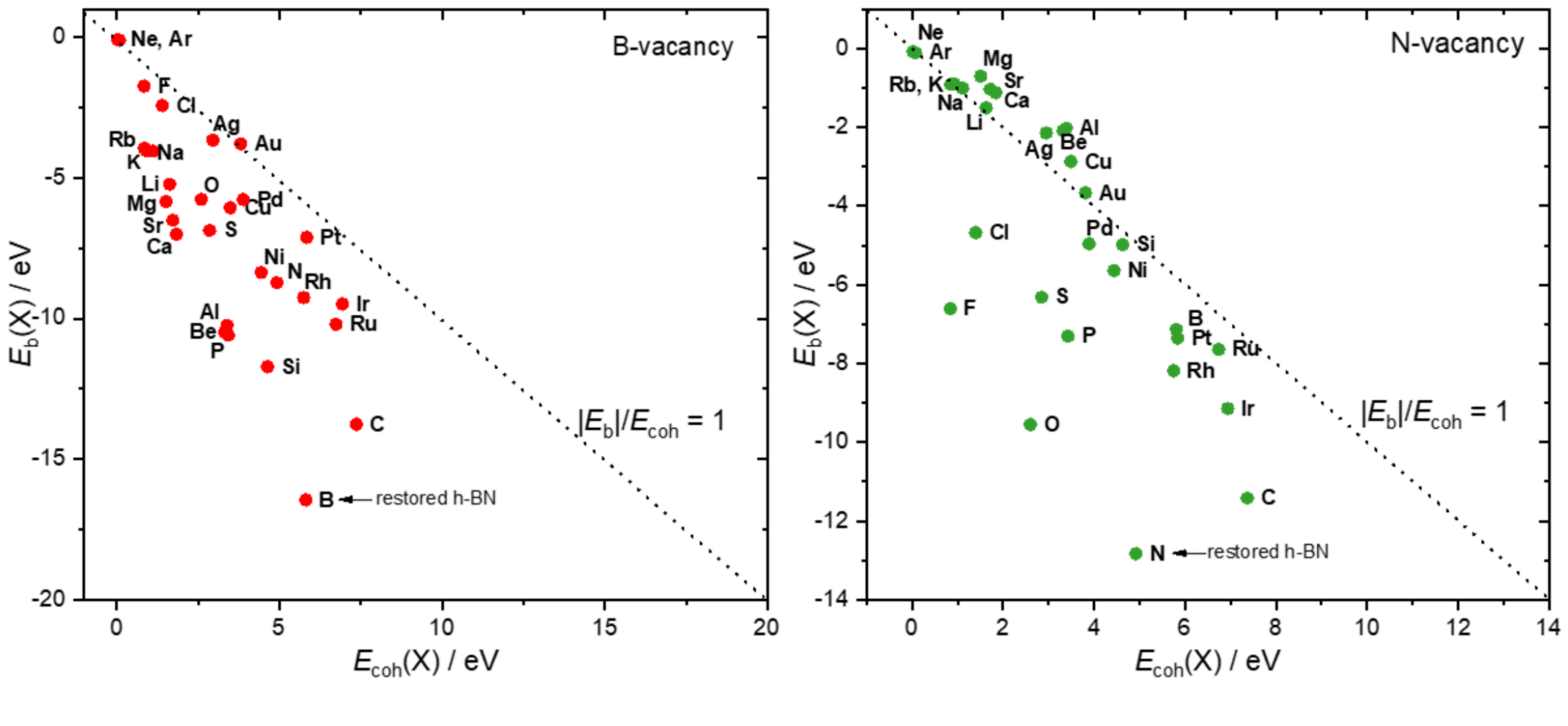

Figure 17. Binding energy of element X on the B-vacancy of h-BN *vs.* its cohesive energy, and binding energy of element X on the N-vacancy of h-BN *vs.* its cohesive energy. Dotted line gives $|E_b|/E_{coh} = 1$.

In contrast, noble gases and heavy main-group elements cluster near or below the diagonal, confirming that their adsorption is dominated by weak van der Waals interactions and is thus readily reversible. At the nitrogen vacancy, the overall scaling shifts toward more positive binding energies, reflecting the weaker interaction of most elements with the three undercoordinated boron atoms. Only a limited set of adsorbates, again dominated by C, N, O, and selected late transition metals, exceed their cohesive energies, indicating that vacancy passivation and strong chemisorption are restricted to these reactive species. Taken together, the scaling behavior in **Figure 17** rationalizes the qualitative trends discussed above: adsorption at $V_B$ is primarily limited by the intrinsic cohesive strength of the element, whereas adsorption at $V_N$ is constrained by the electron deficiency of the surrounding boron atoms. These relationships suggest a simple predictive descriptor, $|E_b|/E_{coh}$, for assessing the stability of isolated adatoms on h-BN defects and for identifying which elements are likely to form substitutional dopants rather than mobile surface species or aggregates. This descriptor directly compares the strength of the adatom-substrate interaction with the element's intrinsic cohesive bonding in its condensed phase, thereby capturing the competition between surface anchoring and adatom aggregation. When $|E_b|/E_{coh} > 1$, adsorption at the defect is energetically favored over metal-metal or atom-atom bonding, indicating that isolated adatoms are thermodynamically stable and unlikely to form clusters. Such conditions are met predominantly at boron vacancies for many transition and coinage metals, as well as for light p-block elements, consistent with the strong trapping behavior observed in the present calculations.

Conversely, when $|E_b|/E_{coh} < 1$, cohesive interactions dominate and adatoms are expected to remain mobile on the surface or to aggregate, even in the presence of defects. This regime is characteristic of noble gases, weakly interacting main-group elements, and several s-block metals at nitrogen vacancies, in agreement with their shallow adsorption potentials and large relaxation distances. Intermediate values of $|E_b|/E_{coh} \approx 1$ mark a transition region in which partial vacancy passivation or metastable adsorption may occur, and where kinetic effects or external conditions could determine whether substitutional incorporation or surface diffusion prevails. In practical terms, this descriptor enables rapid screening of candidate elements for single-atom stabilization, substitutional doping, or reversible functionalization of h-BN surfaces, without requiring detailed system-specific analysis.

### 3.6. Implications for functionalization and applications

The systematic trends identified above provide clear guidance for exploiting defective h-BN in functional applications, which are dominated by applications in the energy field [30]. By introducing vacancies, h-BN can be transformed from a chemically inert wide-band-gap insulator into an active and tunable platform whose reactivity and selectivity are governed by defect type. In this context, vacancies act as chemically programmable sites that enable controlled stabilization of otherwise weakly bound species.

In catalysis, the strong affinity of the boron vacancy for transition and coinage metals indicates that $V_B$ sites can serve as effective anchoring centers for single-atom catalysts. Binding energies that surpass the cohesive energies of the corresponding metals (e.g., Ni −8.4 eV, Ru −10.2 eV, Pt −7.1 eV) suggest that isolated metal atoms are thermodynamically stabilized against aggregation, fulfilling a key requirement for single-atom catalysis. Additionally, the significant band-gap reduction and, in some cases, the emergence of nearly metallic electronic structures indicate efficient electronic coupling between the active metal center and the substrate. These features are especially relevant for reactions that benefit from high atomic dispersion and accessible charge transfer, such as hydrogen evolution, $CO_2$ reduction, and nitrogen-containing molecule activation, where h-BN offers structural stabilization without chemically competing with the active site. For sensing and surface functionalization, the nitrogen vacancy offers complementary, more selective reactivity. Its preferential stabilization of electronegative and covalently bonded adsorbates, such as O, F, and Cl, enables targeted modification of the local electronic structure while avoiding the formation of extended metallic conduction pathways. These adsorbates produce localized mid-gap states and significant local dipoles, which are beneficial for chemical sensing, work-function tuning, and interface engineering in electronic and optoelectronic devices. The moderate binding strength of halogens and small non-metals further indicates a balance between stability and reversibility, essential for reusable and responsive sensor materials. Noble gases remain weakly bound on pristine and defective h-BN alike, providing a useful internal benchmark for dispersion-dominated interactions. Their consistently small adsorption energies (≈ −0.05 to −0.1 eV) and negligible impact on the electronic structure make them valuable reference systems for validating van der Waals treatments and for distinguishing genuine chemisorption from numerical artifacts in computational studies.

These insights also extend naturally to heterostructures and composite materials. In layered systems combining h-BN with graphene or transition-metal dichalcogenides, controlled vacancy formation can define chemically active interfacial junctions while preserving the mechanical integrity and dielectric properties of h-BN. Likewise, vacancy-stabilized metal atoms incorporated into h-BN-based composites may enable hybrid materials with tailored catalytic, electronic, or sensing functionalities. Overall, the defect-dependent trends established here outline a general strategy for rational functionalization of boron nitride surfaces, in which vacancy type and concentration, together with the choice of adsorbate, serve as tunable parameters for balancing stability, electronic structure, and chemical selectivity.

## 4. Conclusions

A comprehensive density functional theory investigation has been carried out to elucidate adsorption trends across the periodic table on pristine and defected hexagonal boron nitride. The results demonstrate that the intrinsic chemical inertness of pristine h-BN can be fundamentally altered by the introduction of atomic vacancies. Whereas adsorption on the pristine surface is generally weak and dominated by dispersion interactions, the formation of boron and nitrogen vacancies creates localized, chemically active sites that enable strong and selective chemisorption. The two vacancy types exhibit distinct and complementary reactivity. The boron vacancy, surrounded by electron-rich nitrogen atoms, functions as a strongly electron-donating site that stabilizes metallic and electropositive adsorbates, frequently with binding energies exceeding the cohesive energies of the corresponding elemental phases. This behavior indicates a thermodynamic preference for isolated adatoms over clustering and identifies $V_B$ as an effective anchoring center for transition metals and a promising platform for single-atom catalysis. In contrast, the nitrogen vacancy, coordinated by electron-deficient boron atoms, preferentially stabilizes electronegative and covalently bonding species such as C, N, O, and F, whereas it interacts more weakly with metallic adsorbates. This selectivity enables targeted functionalization without inducing extensive metallic conduction.

A systematic comparison of adsorption and cohesive energies establishes a simple, predictive descriptor for adsorption stability on defective h-BN, delineating regimes of physisorption, strong chemisorption, and vacancy passivation. The associated electronic-structure modifications span a broad range, from localized mid-gap states to near-metallic behavior, depending on the adsorbate and the vacancy type. Overall, this work provides a unified and physically transparent framework for understanding adsorption on h-BN across the periodic table. By explicitly linking vacancy coordination, adsorption strength, and electronic response, it highlights vacancy engineering as a powerful strategy for tailoring the catalytic, sensing, and electronic functionalities of boron nitride–based materials.

## CRediT authorship contribution statement

**Ana S. Dobrota**: Formal analysis, Investigation, Data curation, Writing - original draft, Visualization. Formal analysis, Investigation. **Natalia V. Skorodumova**: Resources, Writing - review & editing, Supervision, Project administration, Funding acquisition. **Igor A. Pašti**: Conceptualization, Methodology, Validation, Resources, Writing - original draft, Supervision, Writing - review & editing, Funding acquisition.

## Declaration of Competing Interest

The authors declare that they have no known competing financial interests or personal relationships that could have appeared to influence the work reported in this paper.

## Acknowledgement

A.S.D. and I.A.P. acknowledge the financial support provided by the Serbian Ministry of Science, Technological Development, and Innovations (contract no. 451-03-137/2025-03/200146) and the Serbian Academy of Sciences and Arts (projects F-49 and F-190). The computations and data handling were enabled by resources provided by the National Academic Infrastructure for Supercomputing in Sweden (NAISS) at the National Supercomputer center (NSC) at Linköping University, partially funded by the Swedish Research Council through grant agreement No. NAISS 2024/5-718.

## Declaration of generative AI and AI-assisted technologies in the writing process

During the preparation of this work, the authors used ChatGPT in order to improve readability and language. After using this tool, the authors reviewed and edited the content as needed and take full responsibility for the content of the publication.

## Data Availability Statement

Data are available upon reasonable request.

# Periodic Trends in Single−Atom Adsorption on h−BN: From Pristine Surfaces to Vacancy−Engineered Sites

Ana S. Dobrota[1], Natalia V. Skorodumova[2], Igor A. Pašti[1,3]*
[1] *University of Belgrade – Faculty of Physical Chemistry, Studentski trg 12−16, 11000 Belgrade, Serbia*
[2] Applied Physics, Division of Materials Science, Department of Engineering Sciences and Mathematics, Luleå University of Technology, 971 87 Luleå, Sweden
[3] Serbian Academy of Sciences and Arts, Kneza Mihaila 35, 11000 Belgrade, Serbia

*corresponding author
Prof. Igor A. Pašti
*University of Belgrade – Faculty of Physical Chemistry*
*Studentski trg 12−16, 11158 Belgrade, Serbia*
*E−mail: igor@ffh.bg.ac.rs*
*Phone: +381 11 3336 625*

## S.1. Hexagonal boron-nitride: adsorption sites

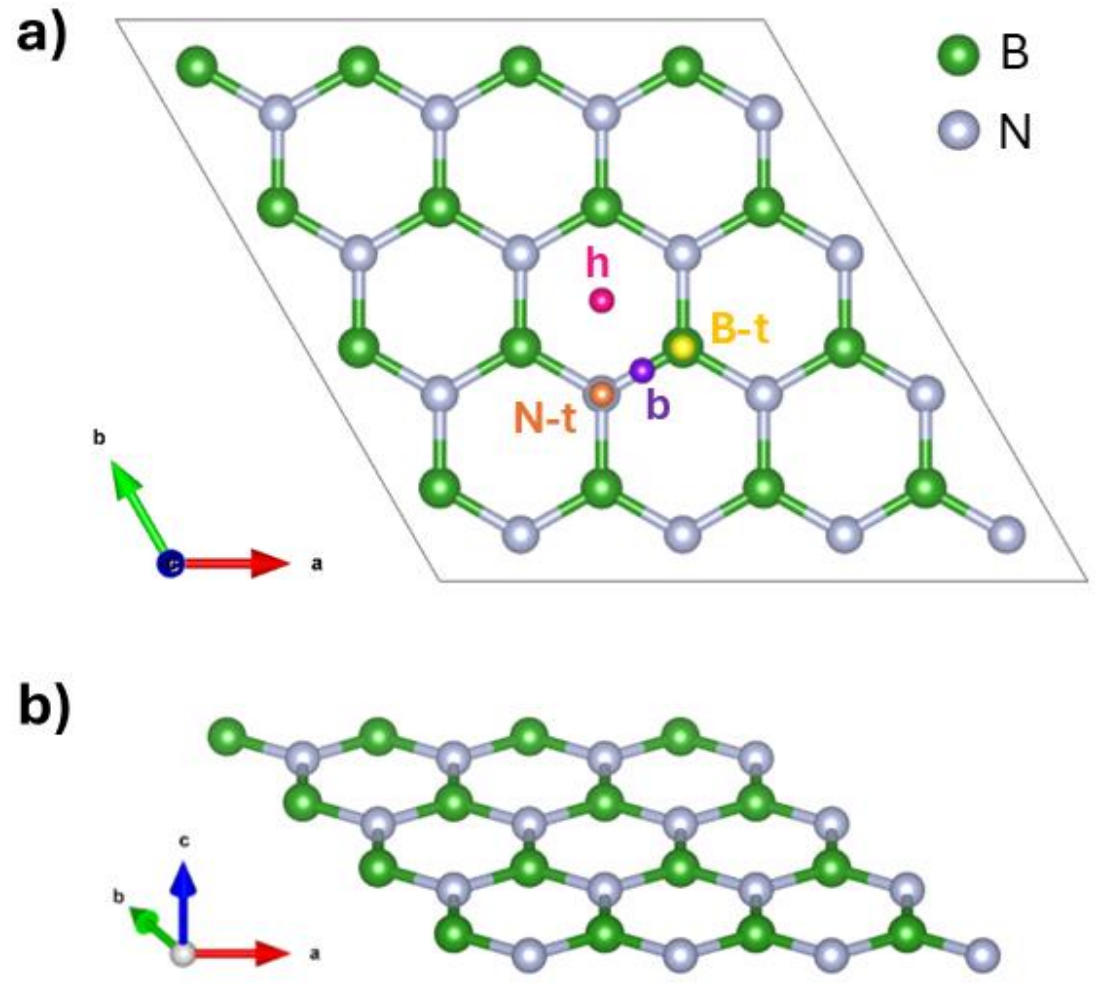

**Figure S1.** Left: the optimized structure of 2D *h*-BN: a) top view with supercell outlines, and investigated adsorption sites: on top of N (N-t, N-top), on top of B (B-t, B-top), on B−N bridge (b), and at the center of $B_3N_3$ ring (hollow - h); b) side view.

**Table S1.** Binding energies of selected elements ($E_b$(A)) and magnetic moments ($M$) on different adsorption sites on pristine h-BN.

| **Element** | **Group** | **Period** | **Site** | $E_b$**(A) / eV** | $M$ **/ $\mu_B$** |
|---|---|---|---|---|---|
| H | 1 | 1 | bridge | −0.05 | 1.0 |
| H | 1 | 1 | B−top | −0.02 | 1.0 |
| H | 1 | 1 | hollow | −0.03 | 1.0 |
| H | 1 | 1 | N−top | −0.04 | 1.0 |
| Li | 1 | 2 | bridge | −0.31 | 1.0 |
| Li | 1 | 2 | B−top | 0.02 | 1.0 |
| Li | 1 | 2 | hollow | −0.08 | 1.0 |
| Li | 1 | 2 | N−top | −0.12 | 1.0 |
| Na | 1 | 3 | bridge | −0.21 | 1.0 |
| Na | 1 | 3 | B−top | −0.2 | 1.0 |
| Na | 1 | 3 | hollow | −0.2 | 1.0 |
| Na | 1 | 3 | N−top | −0.18 | 1.0 |
| K | 1 | 4 | bridge | −0.21 | 1.0 |
| K | 1 | 4 | B−top | −0.05 | 1.0 |
| K | 1 | 4 | hollow | −0.07 | 1.0 |
| K | 1 | 4 | N−top | −0.21 | 1.0 |
| Rb | 1 | 5 | bridge | −0.07 | 1.0 |
| Rb | 1 | 5 | B−top | −0.21 | 1.0 |
| Rb | 1 | 5 | hollow | −0.07 | 1.0 |
| Rb | 1 | 5 | N−top | −0.21 | 1.0 |
| Be | 2 | 2 | bridge | −0.2 | 0.0 |
| Be | 2 | 2 | B−top | −0.2 | 0.0 |
| Be | 2 | 2 | hollow | 1.73 | 1.98 |
| Be | 2 | 2 | N−top | −0.18 | 0.0 |
| Mg | 2 | 3 | bridge | −0.24 | 0.0 |
| Mg | 2 | 3 | B−top | −0.23 | 0.0 |
| Mg | 2 | 3 | hollow | −0.25 | 0.0 |
| Mg | 2 | 3 | N−top | −0.24 | 0.0 |
| Ca | 2 | 4 | bridge | −0.31 | 0.0 |
| Ca | 2 | 4 | B−top | −0.3 | 0.0 |
| Ca | 2 | 4 | hollow | −0.31 | 0.0 |
| Ca | 2 | 4 | N−top | −0.31 | 0.0 |
| Sr | 2 | 5 | bridge | −0.32 | 0.0 |
| Sr | 2 | 5 | B−top | −0.31 | 0.0 |
| Sr | 2 | 5 | hollow | −0.33 | 0.0 |
| Sr | 2 | 5 | N−top | −0.32 | 0.0 |
| Ru | 8 | 5 | bridge | −1.78 | 2.0 |
| Ru | 8 | 5 | B−top | −1.65 | −2.0 |
| Ru | 8 | 5 | hollow | −2.19 | 2.0 |
| Ru | 8 | 5 | N−top | −1.66 | −2.0 |
| Rh | 9 | 5 | bridge | −1.65 | 1.0 |
| Rh | 9 | 5 | B−top | −1.41 | −1.0 |
| Rh | 9 | 5 | hollow | −1.62 | 1.0 |
| Rh | 9 | 5 | N−top | 12.4 | 0.77 |
| Ir | 9 | 6 | bridge | −1.3 | 1.0 |
| Ir | 9 | 6 | B−top | −1.02 | −3.0 |
| Ir | 9 | 6 | hollow | −1.16 | 3.0 |
| Ir | 9 | 6 | N−top | −1.29 | 1.0 |
| Ni | 10 | 4 | bridge | −1.37 | 0.0 |
| Ni | 10 | 4 | B−top | −0.65 | 0.0 |
| Ni | 10 | 4 | hollow | −1.16 | 0.0 |
| Ni | 10 | 4 | N−top | −1.57 | 0.0 |
| Pd | 10 | 5 | bridge | −1.12 | 0.0 |
| Pd | 10 | 5 | B−top | −0.98 | 0.0 |
| Pd | 10 | 5 | hollow | −0.95 | 0.0 |
| Pd | 10 | 5 | N−top | N.B. | −3.7 |
| Pt | 10 | 6 | bridge | −1.63 | 0.0 |
| Pt | 10 | 6 | B−top | −0.85 | 0.0 |
| Pt | 10 | 6 | hollow | −0.9 | 0.0 |
| Pt | 10 | 6 | N−top | −1.8 | 0.0 |
| Cu | 11 | 4 | bridge | −0.34 | 1.0 |
| Cu | 11 | 4 | B−top | −0.34 | 1.0 |
| Cu | 11 | 4 | hollow | −0.26 | 1.0 |
| Cu | 11 | 4 | N−top | −0.39 | 1.0 |
| Ag | 11 | 5 | bridge | −0.31 | −1.0 |
| Ag | 11 | 5 | B−top | −0.29 | 1.0 |
| Ag | 11 | 5 | hollow | −0.29 | 1.0 |
| Ag | 11 | 5 | N−top | −0.31 | −1.0 |
| Au | 11 | 6 | bridge | −0.42 | −1.0 |
| Au | 11 | 6 | B−top | −0.42 | 1.0 |
| Au | 11 | 6 | hollow | −0.37 | 1.0 |
| Au | 11 | 6 | N−top | −0.43 | 1.0 |
| B | 13 | 2 | bridge | −0.85 | 1.0 |
| B | 13 | 2 | B−top | −0.49 | 1.0 |
| B | 13 | 2 | hollow | −0.26 | 1.0 |
| B | 13 | 2 | N−top | −0.54 | 1.0 |
| Al | 13 | 3 | bridge | −0.27 | 1.0 |
| Al | 13 | 3 | B−top | 0.9 | −1.0 |
| Al | 13 | 3 | hollow | −0.13 | 1.0 |
| Al | 13 | 3 | N−top | −0.21 | 1.0 |
| C | 14 | 2 | bridge | −1.09 | 0.0 |
| C | 14 | 2 | B−top | −0.2 | 2.0 |
| C | 14 | 2 | hollow | −0.32 | 2.0 |
| C | 14 | 2 | N−top | −1.31 | 2.0 |
| Si | 14 | 3 | bridge | −0.78 | 2.0 |
| Si | 14 | 3 | B−top | −0.44 | 2.0 |
| Si | 14 | 3 | hollow | −0.45 | 2.0 |
| Si | 14 | 3 | N−top | −0.94 | 2.0 |
| N | 15 | 2 | bridge | −0.57 | 1.0 |
| N | 15 | 2 | B−top | −0.08 | 3.0 |
| N | 15 | 2 | hollow | −0.06 | 3.0 |
| N | 15 | 2 | N−top | −0.29 | 1.0 |
| P | 15 | 3 | bridge | 0.04 | 1.0 |
| P | 15 | 3 | B−top | −0.11 | 3.0 |
| P | 15 | 3 | hollow | −0.11 | 3.0 |
| P | 15 | 3 | N−top | 0.23 | 1.0 |
| O | 16 | 2 | bridge | −2.08 | 0.0 |
| O | 16 | 2 | B−top | −1.09 | 2.0 |
| O | 16 | 2 | hollow | −0.29 | 2.0 |
| O | 16 | 2 | N−top | −1.58 | 0.0 |
| S | 16 | 3 | bridge | −0.94 | 0.0 |
| S | 16 | 3 | B−top | −0.25 | −2.0 |
| S | 16 | 3 | hollow | 0.52 | 0.0 |
| S | 16 | 3 | N−top | −1.06 | 0.0 |
| F | 17 | 2 | bridge | −1.69 | 0.85 |
| F | 17 | 2 | B−top | −1.94 | 0.25 |
| F | 17 | 2 | hollow | −0.82 | 1.0 |
| F | 17 | 2 | N−top | −1.21 | 1.0 |
| Cl | 17 | 3 | bridge | −0.7 | 1.0 |
| Cl | 17 | 3 | B−top | −0.54 | 1.0 |
| Cl | 17 | 3 | hollow | −0.52 | 0.88 |
| Cl | 17 | 3 | N−top | −0.72 | 1.0 |
| He | 18 | 1 | bridge | −0.02 | 0.0 |
| He | 18 | 1 | B−top | 0.0 | 0.0 |
| He | 18 | 1 | hollow | −0.01 | 0.0 |
| He | 18 | 1 | N−top | 0.0 | 0.0 |
| Ne | 18 | 2 | bridge | −0.04 | 0.0 |
| Ne | 18 | 2 | B−top | −0.04 | 0.0 |
| Ne | 18 | 2 | hollow | −0.03 | 0.0 |
| Ne | 18 | 2 | N−top | −0.04 | 0.0 |
| Ar | 18 | 3 | bridge | −0.06 | 0.0 |
| Ar | 18 | 3 | B−top | −0.05 | 0.0 |
| Ar | 18 | 3 | hollow | −0.08 | 0.0 |
| Ar | 18 | 3 | N−top | −0.05 | 0.0 |

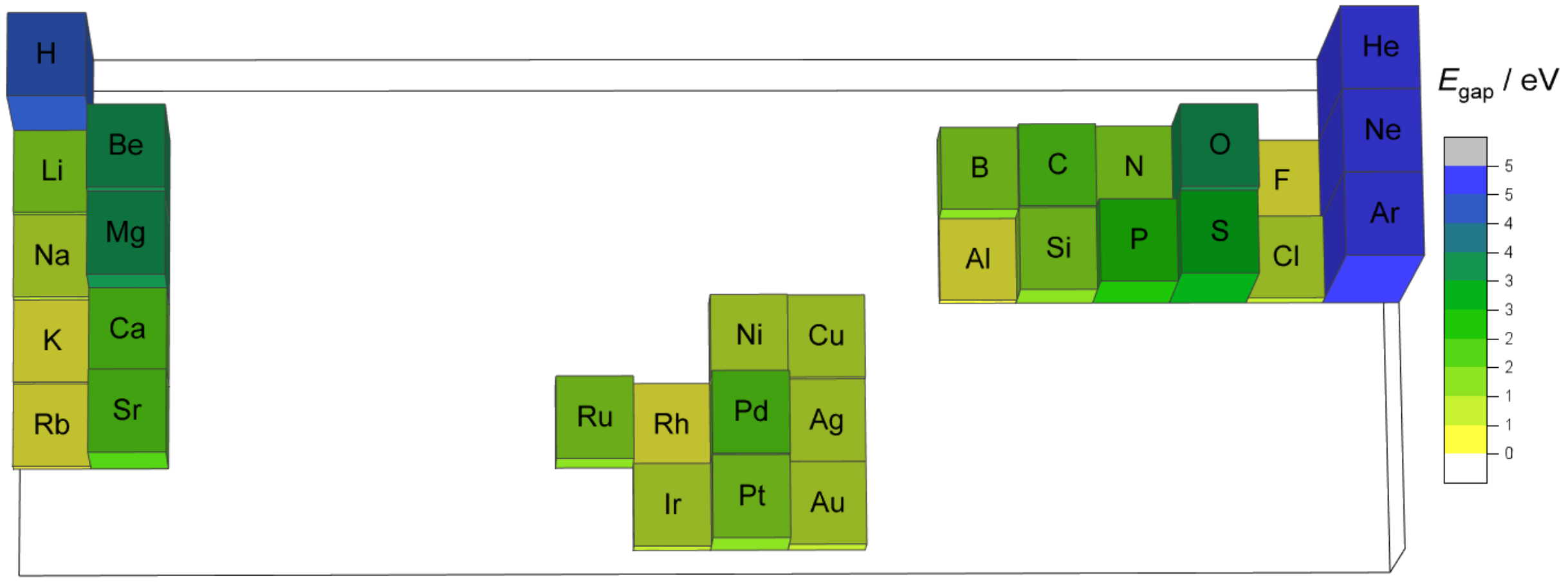

**Figure S2.** Band gaps (in eV) for h-BN with atoms adsorbed on preferential sites

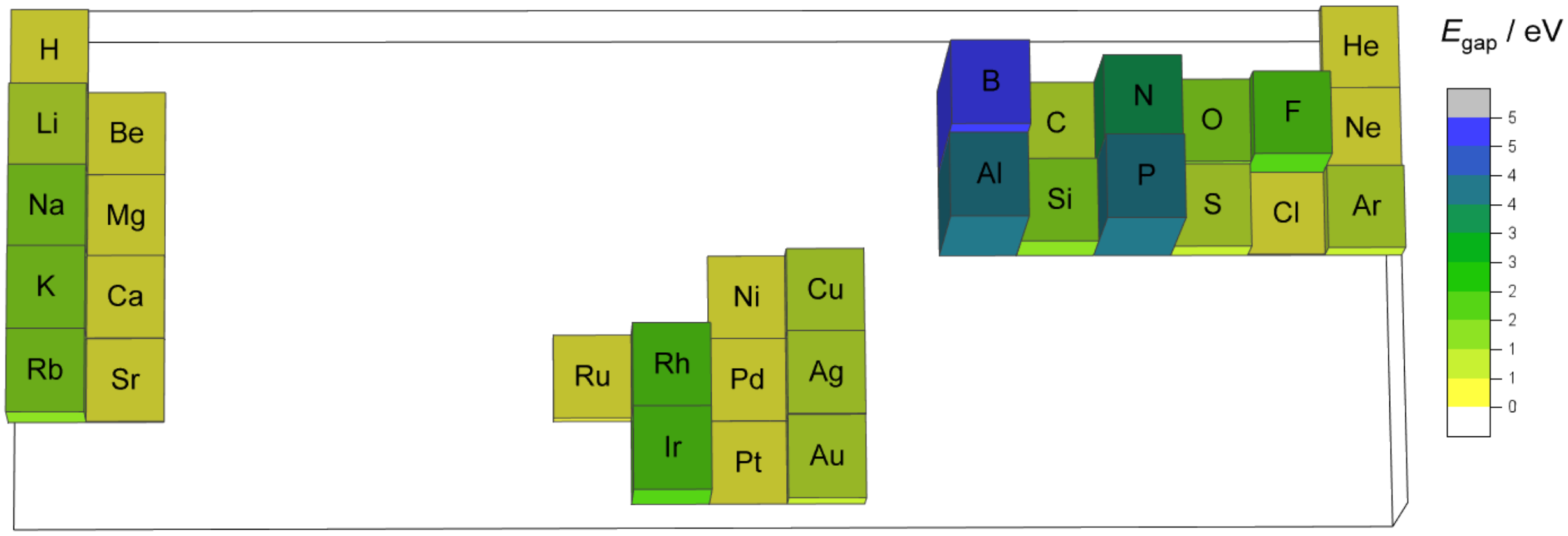

**Figure S3.** Band gaps (in eV) for h-BN with atoms adsorbed on the boron vacancy site

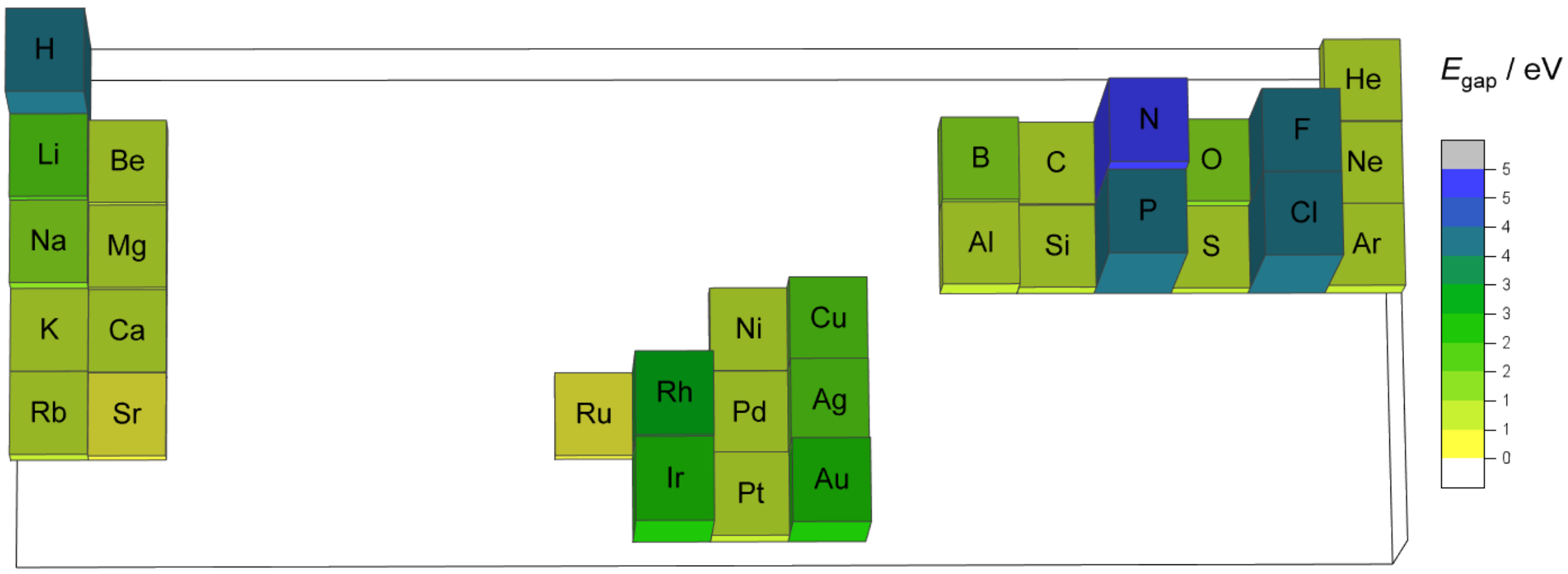

**Figure S4.** Band gaps (in eV) for h-BN with atoms adsorbed on the nitrogen vacancy site